\documentclass{pasj01}
\usepackage{url}
\usepackage{pdflscape}
\usepackage{threeparttable}
\Received{$\langle$reception date$\rangle$}
\Accepted{$\langle$acception date$\rangle$}
\Published{$\langle$publication date$\rangle$}

\RequirePackage{color}

\usepackage[switch,mathlines]{lineno}

\begin{document}

\title{Final Results of Search for New Milky Way Satellites in the Hyper Suprime-Cam Subaru Strategic Program Survey: Discovery of Two More Candidates}

\author{Daisuke~Homma\altaffilmark{1}, Masashi~Chiba\altaffilmark{2}, 
Yutaka~Komiyama\altaffilmark{3}, Masayuki~Tanaka\altaffilmark{1,4}, 
Sakurako~Okamoto\altaffilmark{1,4,5}, Mikito~Tanaka\altaffilmark{3},
Miho~N.~Ishigaki\altaffilmark{1,5}, Kohei~Hayashi\altaffilmark{6},
Nobuo~Arimoto\altaffilmark{5},
Robert~H.~Lupton\altaffilmark{7}, Michael~A.~Strauss\altaffilmark{7}, 
Satoshi~Miyazaki\altaffilmark{1,5}, Shiang-Yu~Wang\altaffilmark{8}, and
Hitoshi~Murayama\altaffilmark{9}
}

\altaffiltext{1}{National Astronomical Observatory of Japan, 2-21-1 Osawa, Mitaka, Tokyo 181-8588, Japan}
\altaffiltext{2}{Astronomical Institute, Tohoku University, Aoba-ku, Sendai 980-8578, Japan}
\altaffiltext{3}{Department of Advanced Sciences, Faculty of Science and Engineering, Hosei University, 184-8584 Tokyo, Japan}
\altaffiltext{4}{The Graduate University for Advanced Studies, Osawa 2-21-1, Mitaka, Tokyo 181-8588, Japan}
\altaffiltext{5}{Subaru Telescope, National Astronomical Observatory of Japan, 650 North A'ohoku Place, Hilo, HI 96720, USA}
\altaffiltext{6}{National Institute of Technology, Sendai College, Natori, Miyagi 981-1239, Japan}
\altaffiltext{7}{Princeton University Observatory, Peyton Hall, Princeton, NJ 08544, USA}
\altaffiltext{8}{Institute of Astronomy and Astrophysics, Academia Sinica, Taipei, 10617, Taiwan}
\altaffiltext{9}{Kavli Institute for the Physics and Mathematics of the Universe (WPI), The University of Tokyo, Kashiwa, Chiba 277-8583, Japan}

\email{d.homma@astr.tohoku.ac.jp}


\maketitle

\begin{abstract}
We present the final results of our search for new Milky Way (MW) satellites using the data from the Hyper Suprime-Cam (HSC) Subaru Strategic Program (SSP) survey over $\sim 1,140$ deg$^2$. In addition to three candidates that we already reported, we have identified two new MW satellite candidates in the constellation of Sextans at a heliocentric distance of $D_{\odot} \simeq 126$~kpc, and Virgo at $D_{\odot} \simeq 151$~kpc, named Sextans~II and Virgo~III, respectively. Their luminosities (Sext~II:$M_V\simeq-3.9$~mag; Vir~III:$M_V\simeq-2.7$~mag) and half-light radii (Sext~II:$r_h\simeq154$~pc; Vir~III:$r_h\simeq 44$~pc) place them in the region of size-luminosity space of ultra-faint dwarf galaxies (UFDs). Including four previously known satellites, there are a total of nine satellites in the HSC-SSP footprint. This discovery rate of UFDs is much higher than that predicted from the recent models for the expected population of MW satellites in the framework of cold dark matter models, thereby suggesting that we encounter a too many satellites problem. Possible solutions to settle this tension are also discussed.
\end{abstract}

\section{Introduction}

The nature of dark matter in the Universe is still enigmatic (e.g., \cite{Bullock2017,Ferreira2021}). For many years, cold dark matter models in a $\Lambda$-dominated Universe ($\Lambda$CDM) have been the theoretical paradigm, because the models are successful for reproducing the observed large-scale structure of the Universe, at scales larger than $\sim 1$~Mpc (e.g., \cite{Tegmark2004,Planck2020}). The theory is based on the hierarchical assembly of smaller to larger dark halos, because the amplitudes of the initial density fluctuations are larger at smaller scales. Thus, smaller halos have higher internal densities and so were able to collapse earlier in the expansion of the Universe. Merging and accretion of these smaller halos in a hierarchical manner form more massive, larger halos as well as larger cosmic structures, which are found to be in agreement with various observations of structures in the Universe.

However, this assembly process leads to the survival of dense parts of the dark matter distribution, giving rise to hundreds of small subhalos surviving in a Milky Way (MW)-sized host halo, in apparent contradiction with the observed population of only tens of MW satellites (see, e.g., \cite{Bullock2017,Drlica-Wagner2020}) --- some of them are known as "classical" dwarfs with $V$-band absolute magnitude, $M_V$, in the range $-18$~mag $\leq$ $M_V$ $\leq$ $-8$~mag, which were discovered before the Sloan Digital Sky Survey (SDSS) \citep{York2000}. This is called the missing satellites problem \citep{Klypin1999,Moore1999}, which has been a key challenge to keep $\Lambda$CDM models. Also, denser smaller halos, which collapsed earlier, tend to sit in the central parts of a larger halo after merging takes place, producing a cuspy central density, whereas some or many of observed dwarf spiral and spheroidal galaxies show a core-like central density in their dark matter distribution \citep{Oh2011, Hayashi2020}. This is called the core/cusp problem. Due to these small-scale issues in $\Lambda$CDM models, say at scales smaller than $\sim 1$~Mpc, several alternative dark matter models, which suppress the amplitude of the power spectrum at smaller scales, have been proposed, such as Warm Dark Matter (WDM), Fuzzy Dark Matter (FDM), and Self-Interacting Dark Matter (SIDM) (e.g., \cite{Spergel2000,Bullock2017,Hui2017,Ferreira2021,Chan2021,Kondo2022}). 

Among these small-scale issues in $\Lambda$CDM models, this work attempts to provide new clues to understand the missing satellites problem. Namely, there is still a possibility that we are undercounting the true population of MW satellites, especially ultra-faint dwarfs (UFDs) with $V$-band absolute magnitude, $M_V$, fainter than $-8$ mag, due to various observational biases related to the survey area and depth \citep{Koposov2008,Tollerud2008}. However, this limitation in our knowledge of MW satellites has been overcome by recent massive imaging surveys, including the SDSS, the Dark Energy Survey (DES) \citep{Abbott2016}, and the Pan-STARRS~1 (PS1) survey \citep{Chambers2016}. Indeed, these modern surveys have discovered a number of new UFDs, giving important insights into the missing satellites problem (e.g., \cite{Willman2005, Sakamoto2006, Belokurov2006a, Laevens2014, Kim2015, KimJerjen2015, Laevens2015a, Laevens2015b, Bechtol2015, Koposov2015, Drlica-Wagner2015, Homma2019}). These survey also imply that faint, diffuse UFDs at larger heliocentric distances, i.e., in the outer parts of the MW halo, still remain undetected and these are actually accessible with Subaru Hyper Suprime-Cam (HSC, \cite{Miyazaki2018}).

The purpose of this paper is to report the final results of our extensive search for UFDs from the HSC Subaru Strategic Program (HSC-SSP) \citep{Aihara2018a,Aihara2018b}. This prime-focus camera, HSC, has a 1.5~deg diameter field of view, and a 330-night survey with this camera has been conducted in the HSC-SSP framework. We have already reported the discovery of three new UFDs: Virgo~I, Cetus~III, and Bo\"otes IV, from the internally released data before 2018 April (S18A) over an area of $\sim 676$~deg$^2$ \citep{Homma2016,Homma2018,Homma2019}. Here, using the data obtained before 2021 January (internal data release S21A), we present the discovery of two more candidate UFDs, Sextans~II and Virgo~III, as well as the identification of seven previously discovered or known satellites. In total, there are nine satellites in the HSC footprint, i.e., five new satellites discovered by our group in addition to four previously known satellites: Sextans, Leo~IV, Leo~V, and Pegasus~III.  This discovery rate sets important constraints on the missing satellites problem in $\Lambda$CDM models.

This paper is organized as follows. The data of HSC-SSP and the method to derive stellar overdensities using isochrone filters are given in Section 2. The results of searching for new satellites in the HSC-SSP footprint are presented in Section 3. Section 4 is devoted to the discussion on the implications for dark matter models, and our conclusions are presented in Section 5. In this paper, magnitudes are given in the AB system \citep{Oke1983}.


\begin{figure*}[t!]
\begin{center}
\includegraphics[width=150mm]{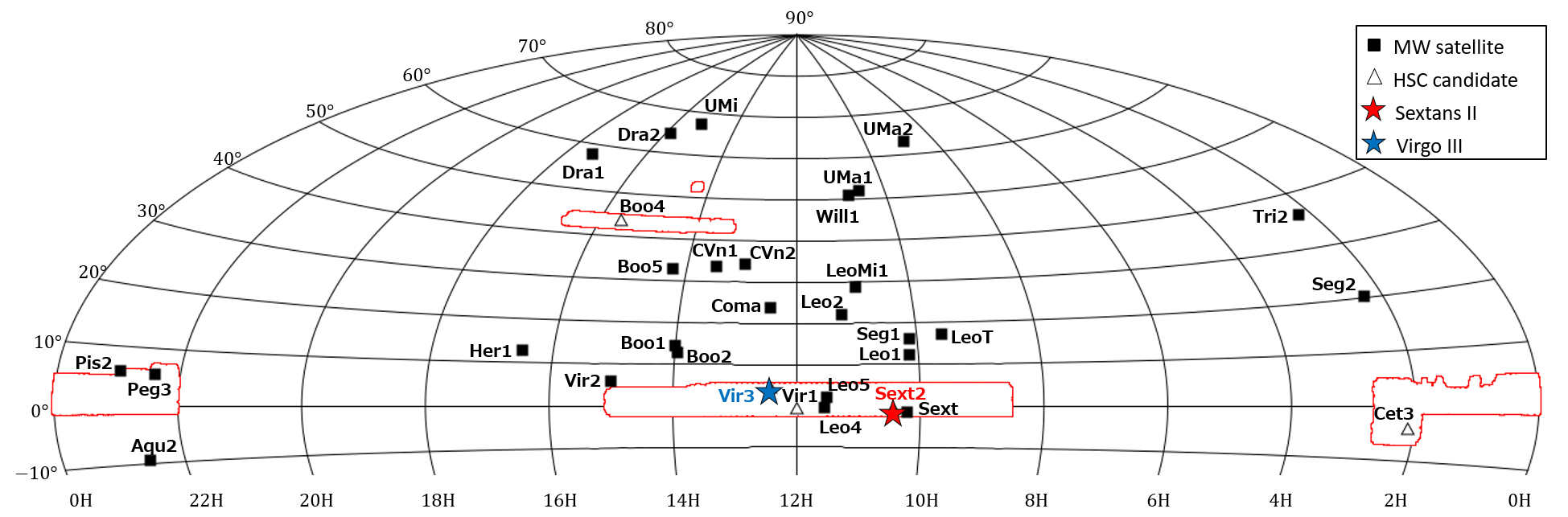}
\end{center}
\caption{
Survey areas of the HSC-SSP Wide layer for the S21A release (bounded by red curves). The red and blue stars denote the locations of the newly found satellites in this work, Sextans II and Virgo III, respectively. The black squares and triangles denote, respectively, the known satellites and those found in the HSC-SSP reported earlier within these areas. We note that Pisces II (shown with Pis2) is just outside the HSC-SSP footprint.
}
\label{fig: survey_area}
\end{figure*}

\section{Data and Method}

The HSC-SSP survey provides multi-band ($g$, $r$, $i$, $z$, $y$) photometry data in the northern sky \citep{Aihara2018a, Aihara2018b, Furusawa2018, Kawanomoto2018, Komiyama2018, Miyazaki2018}. In the Wide layer that we adopt here, the sky coverage is $\sim 1,200$ deg$^2$ and its target limiting magnitudes for a 5$\sigma$ point source are given as ($g$, $r$, $i$, $z$, $y$) = (26.5, 26.1, 25.9, 25.1, 24.4) mag. As in our previous papers, we make use of the $g$, $r$, and $i$-band data obtained before 2021 January (internal data release S21A), which cover $\sim 1,140$ deg$^2$ in two separate fields along the celestial equator and one field around $(\alpha_0, \delta_0)=(242^{\circ},43^{\circ})$ (Figure 1).
These photometric data are processed using hscPipe v8.4 \citep{Bosch2018}, which is a branch of the Legacy Survey of Space and Time (LSST) pipeline (see \cite{Ivezic2019, Axelrod2010, Juric2017})  calibrated against PS1 photometric and astrometric data (e.g., \cite{Tonry2012, Schlafly2012, Magnier2013}). These HSC-SSP data are corrected for the mean foreground extinction in the Milky Way \citep{Schlafly2011}.

\subsection{Selection of target stars}

The method for the selection of target stars is the same as that adopted in our previous work: (1) point-like images are selected to avoid galaxies with extended images, (2) the criterion, $g-r<1.0$, is placed to avoid a foreground contamination from M-type main-sequence stars, and (3) to further separate star candidates against other contaminants, the fiducial $g-r$ vs. $r-i$ relation expected for stars is used. We describe each of these steps in more detail in the following.

First, we select point sources based on the {\it extendedness} parameter calculated by the pipeline, which is defined in terms of the ratio of PSF to cmodel fluxes \citep{Abazajian2004}, $f_{\rm PSF}/f_{\rm cmodel}$: an object with $f_{\rm PSF}/f_{\rm cmodel} > 0.985$ is regarded as a point source ({\it extendedness}$=0$). Here, we adopt this parameter measured in the $i$-band, since the typical seeing in this band is the best of the five filters, with a median of $\sim0.6''$. Also, we decide to select point sources down to $i=24.5$~mag as stars, because the completeness of stars in the star/galaxy separation is larger than 90\% at $i < 22.5$~mag, and decreases to about 50\% at $i=24.5$~mag, based on comparing the HSC data in the COSMOS field with that from HST/ACS \citep{Aihara2018b}.

We then set the color criteria to avoid foreground disk stars and other contaminants, i.e., background quasars and distant galaxies, because these contaminants may often appear as point sources. For this purpose, we utilize the field called WIDE12H at (RA, DEC) $=(180^\circ, 0^\circ)$, where the star/galaxy separation is robust for bright objects with $i<21$ mag, and adopt the $g-r$ vs. $r-i$ color-color diagram for stars with {\it extendedness}$=0$ and galaxies with {\it extendedness}$=1$. As shown in \citet{Homma2018} (Figure 1), stars are distributed in a narrow sequence in this color-color diagram. We first remove M dwarfs in the MW disk characterized by the color $g-r \ge 1$, then set color cuts to select remaining candidate halo stars, with boudaries ($g-r$, $r-i$)$=$(1.00, 0.27), (1.00, 0.57), ($-0.4$, $-0.55$), and ($-0.4$, $-0.25$) and with a width of $\Delta(r-i)=0.3$ mag, which is wider than twice the typical 1$\sigma$ photometric error of $r-i$ at $i=24.5$~mag (see \cite{Homma2018}). We utilize these color cuts to select the MW halo stars in this work.

\subsection{Algorithm for the detection of candidate UFDs}

As in our previous work \citep{Homma2016, Homma2018}, we search for overdensities of stars compared to randomly distributed field stars and distant galaxies/quasars, after filtering stars using the isochrone filter as shown below \citep{Koposov2008, Walsh2009}.

\begin{figure*}[t!]
\begin{center}
\includegraphics[width=150mm]{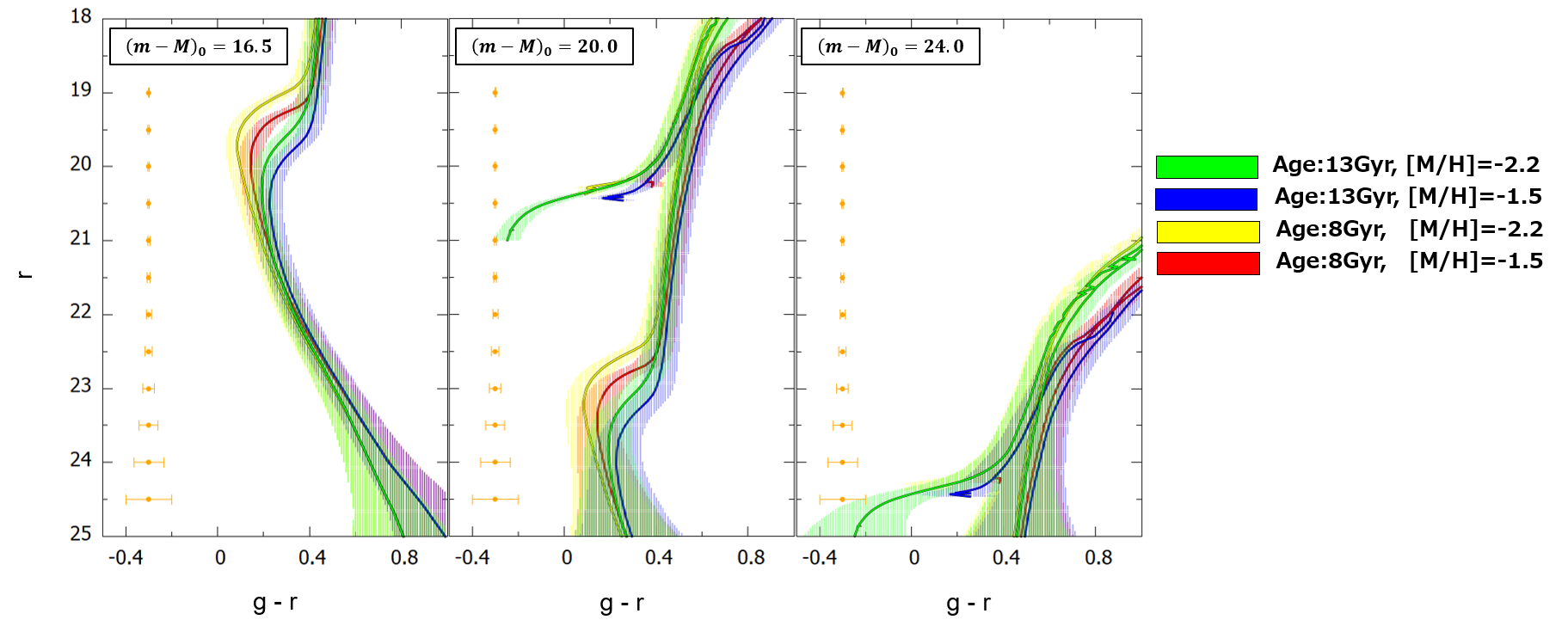}
\end{center}
\caption{
Isochrone filters adopted in this work placed at distance modulus of $(m-M)_0=16.5$, $20.0$, and $24.0$~mag, respectively. Four filters are shown with
(1) an age of $t = 13$~Gyr and metallicity of [M$/$H]$=-2.2$ (green area),
(2) $t= 13$~Gyr and [M$/$H]$=-1.5$ (blue shading), 
(3) $t = 8$~Gyr and [M$/$H]$=-2.2$ (yellow shading), and
(4) $t = 8$~Gyr and [M$/$H]$=-1.5$ (red shading).
The width of each shaded region corresponds to the typical photometric error in color at each $r$ magnitude with an additional 0.05 mag added in quadrature..
}
\label{fig: CMDfilter}
\end{figure*}

\subsubsection{Isochrone filters}
\label{subsec: isochrone}

Our isochrone filters for finding UFDs are designed to select metal-poor, old stars inside a color-magnitude diagram (CMD), since the known UFDs in the MW are characterized by the stellar populations as found in the MW globular clusters. These filters are matched to finding such stars inside a CMD locus within a specified distance range from the Sun. The PARSEC isochrones \citep{Bressan2012} are adopted, with which we use four different models, characterized as 
(a) an age of $t =13$~Gyr and metallicity of [M$/$H]$=-2.2$,
(b) $t =13$~Gyr and [M$/$H]$=-1.5$,
(c) $t = 8$~Gyr and [M$/$H]$=-2.2$, and
(d) $t = 8$~Gyr and [M$/$H]$=-1.5$ (See Fig.~\ref{fig: CMDfilter}).

These filters are defined in a CMD with $g-r$ vs. $r$-band absolute magnitude, $M_r$. The width of each filter, as a function of $r$-band magnitude, is given from the quadrature sum of the 1$\sigma$ error in the HSC photometry of the $g-r$ color and a typical color dispersion ($\sim \pm 0.05$ mag) for red giant-branch stars (RGBs), which corresponds to a typical range of metallicities of $\sim \pm 0.05$ dex within each of the observed MW UFDs. To search for UFDs in the distance range of $D_{\odot} = 20$ to $631$ kpc, we shift the isochrone filter in a CMD over the distance moduli of $(m-M)_0 =16.5$ to $24.0$ mag in steps of $0.5$ mag. Figure 2 shows our four isochrone filters placed at $(m-M)_0 =16.5$, $20.0$, and $24.0$ mag, respectively.

\subsubsection{Method to identify overdensities as UFD candidates}
\label{subsec: overdensities}

Armed with these isochrone filters, we search for statistically significant overdensities as UFD candidates in the following manner.

First, we select and count the stars in $0.05^{\circ} \times 0.05^{\circ}$ boxes in RA and DEC with an overlap of $0.025^{\circ}$ in each direction. This grid interval, $0.05^{\circ}$, in this search is comparable to a half-light diameter of $\sim 80$ pc for a typical UFD located at a distance of $D_{\odot} = 90$ kpc, so that the UFDs at $D_{\odot} \ge 90$ kpc, i.e., our targets in HSC-SSP, can be found within these grids. 

Second, we assemble the number of stars in each cell, $n_{i,j}$ ($i$ for RA, $j$ for DEC), where the cells with no stars, $n_{i,j}=0$, e.g., due to masking near a bright-star image, are ignored, and then estimate, for each of the separate Wide-layer fields, the mean density ($\bar{n}$) and its dispersion ($\sigma$) over all cells. Based on these values, we define the normalized signal in each cell, $S_{i,j}$, which is the number of standard deviations relative to the local mean \citep{Koposov2008, Walsh2009},
$S_{i,j} = (n_{i,j}-\bar{n})/{\sigma}$. We find that this distribution of $S$ is nearly Gaussian, as we found in our previous papers.

Finally, to select overdensities as UFD candidates that have statistically high significance, so that false detections, such as random fluctuations, are avoided, we adopt a detection threshold, $S_{\rm th}$, for the value of $S$ \citep{Walsh2009}. To obtain $S_{\rm th}$, we performed a Monte Carlo analysis as reported in our previous paper \citep{Homma2018}, where for purely random fluctuations in stellar densities, the maximum density contrast, $S_{\rm max}$, as a function of $\bar{n}$ is derived. Then, we adopt a conservative choice for a threshold in the density contrast of $S_{\rm th} = 1.5 \times S_{\rm max}$, with which we identify statistically significant overdensities as UFDs.

\section{Results}
We run the search algorithm explained above on the S21A data of the HSC-SSP. As a result, more than 30 overdensities exceeding the threshold are detected, but about half of them are found to be false due to contamination from artifacts, such as those around bright stars. We distinguish between fake and real overdensities of stars, by looking at the $gri$ color HSC image and the CMD around its location for each of the candidates. The other half of the overdensities are real stellar systems, such as globular clusters(GCs): Pal~3, Whiting 1 \citep{Harris1996}, star cluster(SC): HSC~1 \citep{Homma2019}, and dwarf irregular galaxies(dIrrs): IC1613, Sextans~B, and DDO~190 \citep{McConnachie2012}. The significance and statistical values defined in Section~\ref{subsec: overdensities} of all the overdensities identified as real stellar systems are given in Table~\ref{table: significance}. Note that the dIrrs are located outside the range of $(m-M)_0 \leq24.0$ and not associated with MW, although they are detected by their TRGBs (Tip of the red-giant branch stars) passing through the bottom of the ishochrone filter at $(m-M)_0 \sim16.5$.

In this section, we summarize the nine MW satellites or candidates detected in the HSC-SSP footprint; all of these will be simply referred to as 'MW satellites' in what follows unless otherwise noted. While four of these satellites, Sextans, Leo IV, Leo V, and Pegasus III, are already known, the other five, Virgo I, Cetus III, Bo\"otes IV, Sextans II, and Virgo III, are candidate satellites identified in the HSC-SSP data, among which Sextans II\footnote{Parallel to our study and independently, \citet{Gatto2023} also reported the discovery of Sextans II in the Kilo-Degree Survey \citep{deJong2013}.} and Virgo III are newly found in the S21A data and reported in this paper for the first time.

\begin{table*}[t]
\tbl{Significance of the overdensities identified as stellar systems}{
\begin{tabular}{lcccccccccc}
\hline
Name    &    Classification    &    Filter$^{a}$    &    $(m-M)_0$    &    ($i$, $j$)    &    $n_{i,j}$    &    $\bar{n}$    &    $\sigma$    &    $S_{i,j}$    &    $S_{max}$    &    $S_{i,j}/S_{max}$    \\
    &    &    &    [mag]    &    [deg]    \\
\hline
\hline
Sextans    &    dSph(Classical)    &    (d)    &    $20.0$    &    $(153.250,-1.625)$    &    $157$    &    $1.53$    &    $0.80$    &    $195.7$    &    $7.8$    &    $25.2$  \\
Leo IV    &    dSph(UFD)    &    (a)    &    $21.0$    &    $(173.250,-0.550)$  
  &    $31$    &    $1.84$    &    $1.01$   
 &    $28.9$    &    $7.1$    &    $4.1$  \\
Leo V    &    dSph(UFD)    &    (a)    &    $21.5$    &    $(172.775,2.225)$    &    $38$    &    $1.86$    &    $1.02$    
 &    $35.5$    &    $7.0$    &    $5.1$  \\
Pegasus III    &    dSph(UFD)    &    (a)    &    $21.5$    &   
 $(336.100,5.425)$    &    $25$    &    $2.03$    &    $1.13$    &    $20.3$    &    $6.7$    &    $3.0$  \\
$^{\star}$Virgo I    &    dSph(UFD)    &    (a)    &    $20.0$    &    $(180.025,-0.675)$ 
   &   $16$    &    $1.84$    &    $1.01$    &    $14.0$    &    $7.0$    &    $2.0$  \\
$^{\star}$Cetus III    &    dSph(UFD)    &    (a)    &    $22.0$    &    $(31.327,-4.285)$    &    $13$    &    $1.92$    &    $1.07$    &    $10.4$    &    $6.9$    &    $1.5$  \\
$^{\star}$Bootes IV    &    dSph(UFD)    &    (a)    &    $21.5$    &    $(233.700,43.725)$    &    $26$    &    $1.49$    &    $0.75$    &    $32.9$    &    $7.9$    &    $4.2$  \\
$^{\star}$Sextans II    &    dSph(UFD)    &    (b)    &    $20.0$    &    $(156.425,-0.600)$    &    $21$    &    $1.63$    &    $0.85$    &    $22.7$    &    $7.5$    &    $3.0$  \\
$^{\star}$Virgo III    &    dSph(UFD)    &    (b)    &    $20.5$    &    $(186.350,4.450)$    &    $25$    &    $2.08$    &    $1.15$    &    $19.9$    &    $6.6$    &    $3.0$  \\
Pal 3    &    GC    &    (b)    &    $19.5$    &    $(151.375,0.075)$    &    $322$    &    $1.56$    &    $0.80$    &    $399.5$    &    $7.7$    &    $52.0$  \\
Whiting 1    &    GC    &    (b)    &    $16.5$    &    $(30.750,-3.250)$    &    $139$    &    $2.96$    &    $1.68$    &    $80.8$    &    $6.0$    &    $13.5$  \\
$^{\star}$HSC 1    &    SC    &    (c)    &    $18.0$    &    $(334.325,3.475)$    &    $31$    &    $1.53$    &    $0.79$    &    $37.5$    &    $7.8$    &    $4.8$  \\
IC1613    &    dIrr    &    (b)    &    $24.0$    &    $(16.275,2.025)$    &    $384$    &    $1.48$    &    $0.75$ 
   &    $509.1$    &    $7.9$    &    $64.4$  \\
Sextans B    &    dIrr    &    (d)    &    $16.5$    &    $(149.950,5.300)$    &    $53$    &    $1.64$    &    $0.87$    &    $59.3$    &    $7.5$    &    $7.9$  \\
DDO 190    &    dIrr    &    (d)    &    $17.0$    &    $(216.200,44.525)$    &    $10$    &   $1.30$    &    $0.58$  
  &    $15.1$    &    $8.5$    &    $1.8$  \\
\hline
\end{tabular}}

\begin{tabnote}
$^{a}$ Models of the isochrone filters described in Section~\ref{subsec: isochrone}.
\\
$^{\star}$ Objects discovered in the HSC-SSP survey. Their classifications are not yet uncertain.
\end{tabnote}

\label{table: significance}
\end{table*}

\subsection{Summary of the MW satellites in the HSC-SSP footprint}
We describe below the procedure to calculate the basic properties of these MW satellites, including their heliocentric distances, half-light radii, and absolute magnitudes. These quantities are summarized in Table~\ref{table: list} for all nine satellites identified in the HSC-SSP footprint, where we include the results for four known satellites, Sextans, Leo IV, Leo V, and Peg III, to confirm that our current analysis indeed yields comparable results obtained by previous works. Three candidate satellites that we reported earlier: Vir I, Cet III, and Boo IV, are re-analysed for consistency with the results of our previous analysis \citep{Homma2018,Homma2019}; with the exception of the estimate of V-band absolute magnitudes (see below), all measured properties remain unchanged for these three satellites.

\begin{figure*}[t!]
\begin{center}
\includegraphics[width=120mm]{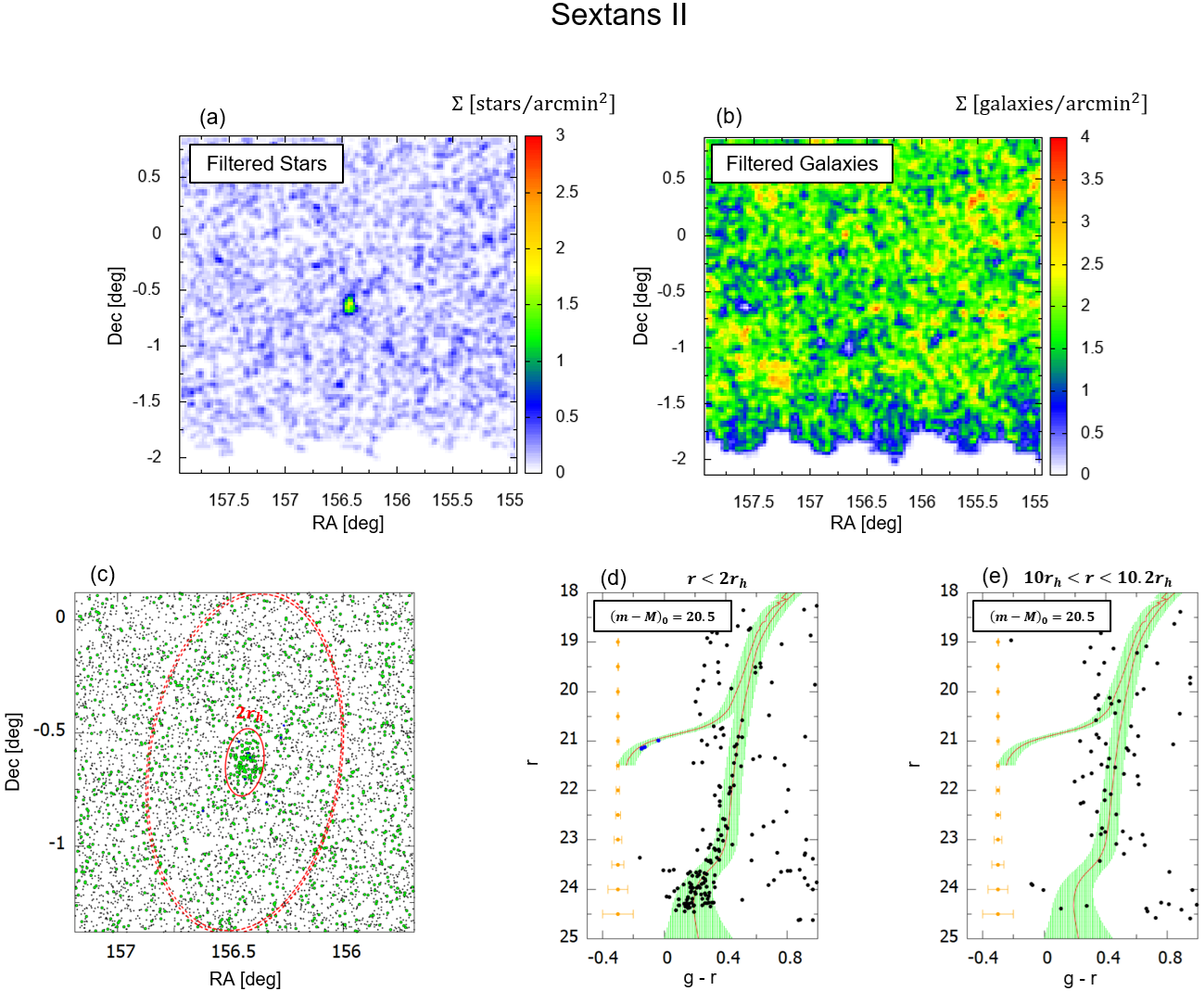}
\end{center}
\caption{
Sextans II. (a) The stellar overdensity passing the isochrone filter of $t=13$ Gyr and [M/H]$=-2.2$ at $(m-M)_0 = 20.5$ for the point sources satisfying $i<24.5$, $g-r<1.0$, and the color-color cuts expected for stars in the $g-r$ vs. $r-i$ diagram (see subsection 2.1). (b) The spatial distribution of the galaxies which pass the same isochrone filter and constraints as for the stars. An overdensity at the center of this plot is seen, possibly because of misclassification of faint stars within a cluster as galaxies in hscPipe. (c) The spatial distribution of the objects classified as stars around the overdensity.  The green circles and black dots denote, respectively, the stars inside and outside the isochrone filter at the distance modulus of $(m-M)_0 = 20.5$. A solid red curve shows an ellipse with a major axis of $r=2.0r_h$ ($r_h=4.2'$) and an ellipticity of $0.43$, whereas dotted red circles show annuli with radii $r=10.0r_h$ and $r=10.2r_h$ from the center of the overdensity. (d) The CMD of the stars in the $g-r$ vs. $r$ relation, which are located within the solid red ellipse in panel (c). Blue circles denote the BHB stars which we will use for an additional distance estimate as described in the text. (e) The same as (d) but for field stars at $10.0r_h < r < 10.2r_h$ having the same solid angle. None of the features of a stellar population (RGBs, MSTOs, BHBs) is apparent in this plot.
}
\label{fig: SextansII_cmd_new}
\end{figure*}

\begin{figure*}[t!]
\begin{center}
\includegraphics[width=120mm]{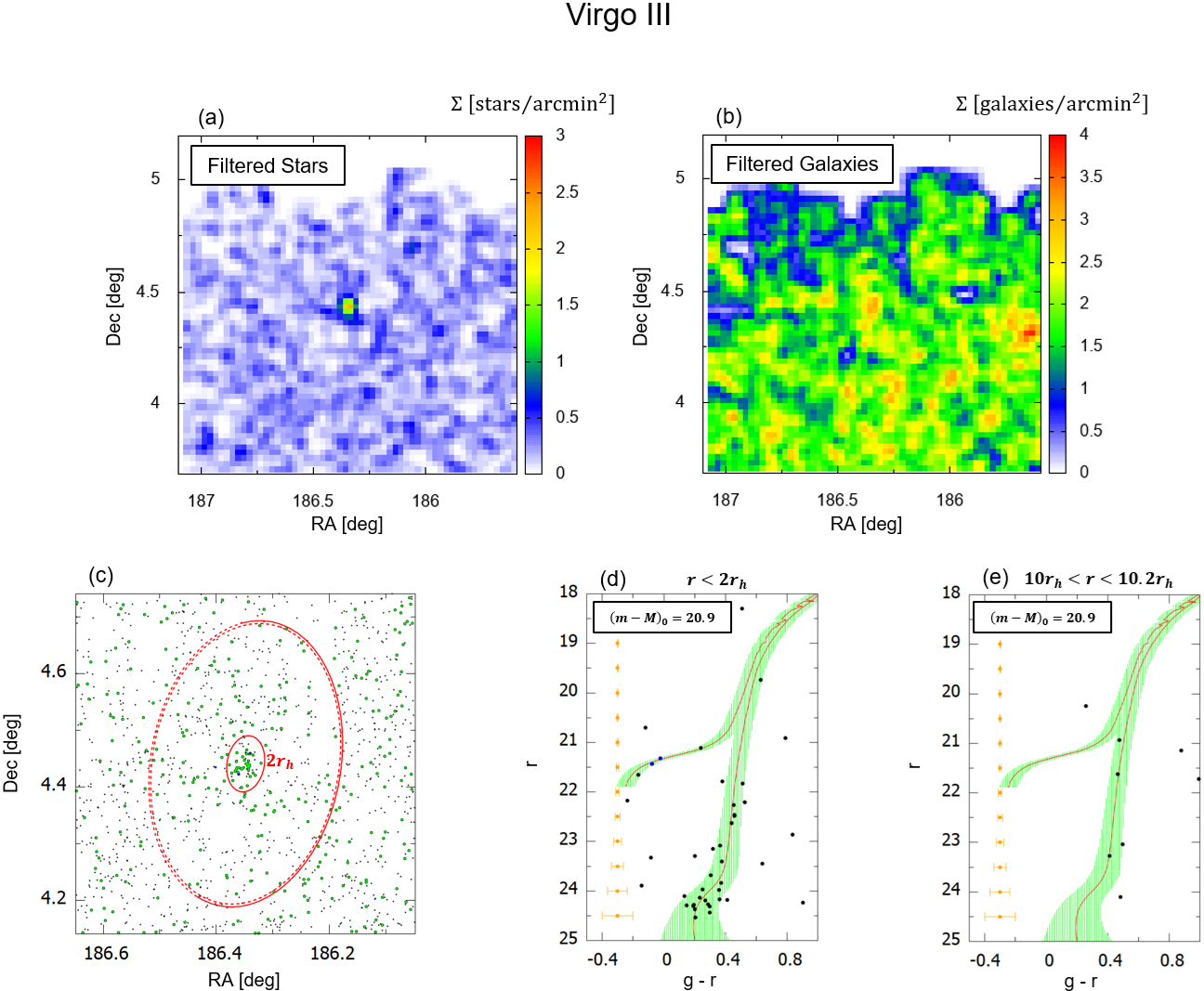}
\end{center}
\caption{
Virgo III. Same as Figure~\ref{fig: SextansII_cmd_new} except that the isochrone filter is at $(m-M)_0 = 20.9$. (c) The solid red curve shows an ellipse with a major axis of $r=2.0r_h$($r_h=1.0'$) and an ellipticity of $0.29$. 
}
\label{fig: VirgoIII_cmd_new}
\end{figure*}

Here, we identify two new candidate MW satellites, which we name Sextans II (Sext II) and Virgo III (Vir III) at (RA, DEC)$=(156.4^\circ, -0.6^\circ)$ and $(186.3^\circ, 4.4^\circ)$, respectively. In Figures \ref{fig: SextansII_cmd_new} and \ref{fig: VirgoIII_cmd_new}, we show the space and CMD distribution of stars in these candidates, demonstrating that they are true stellar systems: RGBs, main sequence turn-off stars (MSTOs), and blue horizontal-branch stars (BHBs) are clearly visible in the CMD. Vir~III is found to have the smaller number of each of RGBs, MSTOs, and BHBs than does Sext~II. These are considered to be UFDs because (1) their distributions in CMDs are qualitatively consistent with old stellar populations and (2) they are spatially extended compared to globular clusters with similar brightness, as we describe in Section 4.1.

For completeness, in Appendix~\ref{sec: appendix_1}, we also show the space and CMD distributions of stars in the other seven satellites (Figures 10-16). We note that the methods to derive the structural parameters (Section~\ref{subsec: structure}) and absolute magnitude (Section~\ref{subsec: abs.magnitude}) of these MW satellites are suitable only for very low luminosity systems such as UFDs, in which the number of detected stars is as small as tens to hundreds. Thus, Sextans, a classical dSph with many more stars detected, is an exception for this analysis, and we only calculate its distance. 

\begin{landscape}
\begin{table}[t]
\begin{threeparttable}[h]
\caption{Properties of previously known and candidate MW satellites in the HSC-SSP footprint}
\label{table: list}
\normalsize
\centering
\begin{tabular}{lcccccccccc} 
    \hline
Name\tnote{a}  &    (R.A., Dec)\tnote{b}    &    ($l$,$b$)\tnote{c}    &    $(m-M)_0$\tnote{d}    &    $D_{\odot}$\tnote{e}    &    $\theta$\tnote{f}    &    $\varepsilon$\tnote{g}     &    $N_{\ast}$\tnote{h}    &    $r_h$\tnote{i}    &    $M_{{\rm tot},V}$\tnote{j}    &    $A_V$\tnote{k}        \\
         &    [deg]    &    [deg]    &    [mag]    &    [kpc]    &    [deg]    &        &        &   [$'$] or [pc]    &    [mag]    &    [mag]    \\
    \hline
    \hline
\multicolumn{11}{c}{\rm } \\
\multicolumn{11}{c}{Previously known MW satellites} \\
\multicolumn{11}{c}{\rm } \\
Sextans    &    $(153.262,-1.615)$    &    $(243.498, 42.272)$    &    $19.8^{+0.2}_{-0.2}$    &    $91^{+9}_{-8}$    &    $-$    &    $-$    &    $-$    &    $-$    &    $-$    &    $0.168$    \\
Leo IV    &    $(173.243,-0.536)$    &    $(265.452,56.515)$    &    $21.0^{+0.2}_{-0.2}$    &    $158^{+16}_{-13}$    &    $116^{+24}_{-19}$    &    $0.17^{+0.08}_{-0.10}$    & $103^{+10}_{-9}$    &    $2.6^{+0.2}_{-0.2}$ or $119^{+25}_{-19}$    &    $-4.50^{+0.35}_{-0.31}$    &    $0.069$    \\
LeoV   &    $(172.784, 2.222)$    &    $(261.851,58.535)$   &    $21.3^{+0.2}_{-0.2}$    &    $182^{+18}_{-16}$    &    $106^{+12}_{-12}$    &   $0.37^{+0.10}_{-0.12}$     &    $50^{+6}_{-6}$    &    $1.2^{+0.2}_{-0.1}$ or $62^{+19}_{-12}$    &    $-3.96^{+0.35}_{-0.34}$    &    $0.073$    \\
Pegasus III    &    $(336.100,5.413)$    &    $(69.851,-41.822)$    &    $21.5^{+0.2}_{-0.2}$    &    $200^{+19}_{-18}$    &    $131^{+14}_{-15}$    &    $0.37^{+0.11}_{-0.13}$    &    $42^{+5}_{-6}$    &    $1.5^{+0.2}_{-0.2}$ or $88^{+22}_{-20}$    &    $-4.11^{+0.42}_{-0.25}$    &    $0.404$    \\
\multicolumn{11}{c}{\rm } \\
    \hline
\multicolumn{11}{c}{\rm } \\
\multicolumn{11}{c}{Candidate MW satellites} \\
\multicolumn{11}{c}{\rm } \\
Virgo I    &    $(180.038,-0.681)$    &    $(276.942,59.578)$    &    $19.8^{+0.2}_{-0.2}$    &    $91^{+9}_{-8}$    &    $62^{+8}_{-13}$    &    $0.59^{+0.12}_{-0.14}$    &    $18^{+5}_{-4}$    &    $1.8^{+0.5}_{-0.4}$ or $47^{+19}_{-13}$    &    $-0.90^{+0.72}_{-0.65}$    &    $0.066$    \\
Cetus III    &    $(31.331,-4.270)$    &    $(163.810,-61.133)$    &    $22.0^{+0.2}_{-0.2}$    &    $251^{+24}_{-22}$    &    $101^{+5}_{-6}$    &    $0.76^{+0.06}_{-0.08}$    & $16^{+3}_{-5}$    &    $1.2^{+0.5}_{-0.2}$ or $90^{+42}_{-17}$    &   $-3.45^{+0.49}_{-0.43}$    &    $0.066$    \\
Bootes IV    &    $(233.689, 43.726)$    &    $(70.682, 53.305)$   &    $21.6^{+0.2}_{-0.2}$    &    $209^{+20}_{-18}$    &    $3^{+4}_{-4}$    &   $0.64^{+0.05}_{-0.05}$     &    $124^{+10}_{-10}$    &    $7.6^{+0.8}_{-0.8}$ or $462^{+98}_{-84}$    &    $-5.34^{+0.29}_{-0.17}$    &    $0.067$    \\
Sextans II    &    $(156.437,-0.631)$    &    $(245.326,45.322)$    &    $20.5^{+0.2}_{-0.2}$    &    $126^{+12}_{-11}$    &    $-16^{+9}_{-9}$    &    $0.43^{+0.07}_{-0.08}$    &    $93^{+10}_{-10}$    &    $4.2^{+0.5}_{-0.5}$ or $154^{+35}_{-30}$    &    $-3.91^{+0.37}_{-0.36}$    &    $0.177$    \\
Virgo III    &    $(186.348,4.441)$    &    $(286.476,66.477)$    &    $20.9^{+0.2}_{-0.2}$    &    $151^{+15}_{-13}$    &    $-24^{+21}_{-26}$    &    $0.29^{+0.15}_{-0.19}$    &    $25^{+5}_{-4}$    &    $1.0^{+0.2}_{-0.2}$ or $44^{+14}_{-12}$    &    $-2.69^{+0.45}_{-0.56}$    &    $0.054$        \\
\multicolumn{11}{c}{\rm } \\
    \hline
\end{tabular}
\begin{tablenotes}
\item[a] Satellite's name
\item[b] Equatorial coordinates (J2000)
\item[c] Galactic coordinates
\item[d] Distance modulus
\item[e] Heliocentric distance calculated as $10^{\frac{(m-M)_0}{5}-2}$ kpc
\item[f] Position angle from north to east
\item[g] Ellipticity defined as $\varepsilon=1-(b/a)$ with $a$ the scale-length of the system along its major axis and $b$ that along its minor axis.
\item[h] Number of detected stars estimated to be members of the stellar system
\item[i] Half light radius
\item[j] V-band absolute magnitude
\item[k] Integrated magnitudes are corrected for the mean Galactic foreground extinction, $A_V$ \citep{Schlafly2011}.
\end{tablenotes}
\end{threeparttable}
\end{table}
\end{landscape}

\subsection{Distance estimate}
\label{subsec: distance}
The first step after identifying the candidate stellar system is to estimate its heliocentric distance, or distance modulus $(m-M)_0$, by fitting the isochrone filter to the overdensity having the highest significance (i.e., $S$ defined in Section~\ref{subsec: overdensities}), where $(m-M)_0$ is shifted in steps of $0.1$ mag around the value derived in detecting it. This evaluation is mostly consistent with another distance estimate below, and we assume that the systematic uncertainty in distance modulus is $0.2$ mag. It is found that for four known satellites, Sextans ($(m-M)_0=19.8 \pm 0.2$ mag from this work), Leo IV ($21.0 \pm 0.2$ mag), Leo V ($21.3 \pm 0.2$ mag), and Peg III ($21.5 \pm 0.2$ mag), our distance estimates are within 1$\sigma$ of previous estimates, $(m-M)_0=19.67 \pm 0.10$, $20.94 \pm 0.09$, $21.25 \pm 0.12$, and $21.56 \pm 0.20$ mag, respectively (\cite{McConnachie2012} for the first three satellites, and \cite{Kim2015} for Peg III). 

The corresponding heliocentric distances to the newly found satellites, Sext II and Vir III, are $126^{+12}_{-11}$~kpc and $151^{+15}_{-13}$~kpc, respectively.

An alternative estimate for the heliocentric distance comes from the BHBs, given their absolute magnitude calibrated by \citet{Deason2011}. The method to select BHBs from the data uses three colors, $g-r$, $r-i$, and $i-z$, as described in \citet{Fukushima2019}. We apply this procedure to the satellites which contain BHBs (except Virgo I and Cetus III), and Table~\ref{table: distance} shows the results. These distances from BHBs are found to be in agreement with those shown in Table~\ref{table: list} with a difference of a few percents.

\begin{table}
\tbl{Distance moduli derived from BHBs}{
\begin{tabular}{lcc}
\hline
Name    &    Number of BHB stars$^{a}$    &    $(m-M)_0$$^{b}$    \\
    &    &    [mag]    \\
\hline
Sextans    &    $195$    &    $19.77\pm{0.20}$ ($19.8\pm{0.20}$)     \\
Leo IV    &    $8$    &    $21.00\pm{0.08}$ ($21.0\pm{0.20}$)     \\
Leo V    &    $8$    &    $21.29\pm{0.05}$ ($21.3\pm{0.20}$)     \\
Pegasus III    &    $5$    &    $21.53\pm{0.06}$ ($21.5\pm{0.20}$)     \\
Bootes IV    &    $10$    &    $21.54\pm{0.19}$ ($21.6\pm{0.20}$)     \\
Sextans II    &    $9$    &    $20.47\pm{0.22}$ ($20.5\pm{0.20}$)     \\
Virgo III    &    $3$    &    $20.91\pm{0.04}$ ($20.9\pm{0.20}$)     \\
\hline
\end{tabular}}
\begin{tabnote}
$^{a}$Within 5 times the half light radius of the centroid.
\\
$^{b}$The values in parentheses are those derived from isochrone shifting, and listed for comparison.
\end{tabnote}
\label{table: distance}
\end{table}

\subsection{Structural parameters}
\label{subsec: structure}
The next step is to calculate the structural properties of each stellar system following \citet{Martin2008} and \citet{Martin2016}. There are six parameters $(\alpha_0, \delta_0, \theta, \epsilon, r_h, N_{\ast})$; $(\alpha_0, \delta_0)$ for the celestial coordinates of the centroid of the overdensity, $\theta$ for its position angle from north to east, $\epsilon$ for the ellipticity, $r_h$ for the half-light radius measured along the major axis, and $N_{\ast}$ for the number of stars within the isochrone and brighter than our magnitude limit, which belong to the overdensity. We then obtain the best-fit parameter set and their uncertainties by performing a maximum likelihood analysis following \citet{Martin2008} for the stars within a circle of $5\sim6$ times the inferred half-light radius. 

We find that for the known UFDs, Leo IV, Leo V and Peg III, for which the above method is applicable, the derived structural parameters are in good agreement with the previous results: Leo~IV: $(r_h({\rm pc}),\varepsilon)=(119^{+25}_{-19},0.17^{+0.08}_{-0.10})$ compared to (127.8, 0.05) by \citet{Sand2010}, Leo~V: $(62^{+19}_{-12},0.37^{+0.10}_{-0.12})$ compared to $(65^{+30}_{-30},0.52)$ by \citet{Sand2012}, and Peg~III: $(88^{+22}_{-20},0.37^{+0.11}_{-0.13})$ compared to $(79^{+30}_{-24},0.46^{+0.18}_{-0.27})$ by \citet{Kim2015}.

Applying this method to Sext~II and Vir~III, we find $(r_h({\rm pc}),\varepsilon)=(154^{+35}_{-30},0.43^{+0.07}_{-0.08})$ and $(44^{+12}_{-12},0.29^{+0.15}_{-0.19})$, respectively.

\begin{figure}[t!]
\begin{center}
\includegraphics[width=85mm]{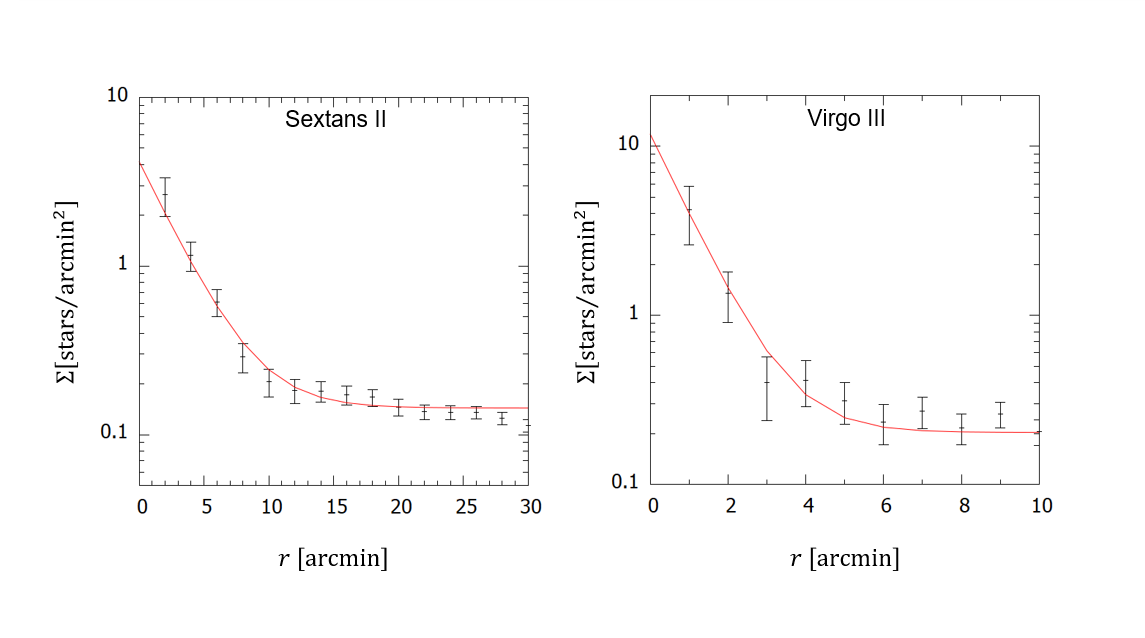}
\end{center}
\caption{
The density distribution of the stars in Sext~II and Vir~III passing the isochrone filter given in Figure 3 and 4, in elliptical annuli as a function of mean radius. The uncertainties are based on Poisson statistics. The red line denotes the best-fit exponential profile within $5\sim6r_h$ plus a constant representing the background.
}
\label{fig: Radial_new}
\end{figure}

In Figure~\ref{fig: Radial_new}, we show the radial distribution of the stars passing the isochrone filter depicted in panel (d) of each of Figures 3 and 4, which is obtained from the count of the average density within elliptical annuli. The red line in the figure denotes the best-fit exponential profile represented by the best-fit parameter set above. The radial distributions of other satellites are shown in Appendix~\ref{sec: appendix_1} (Figure~\ref{fig: Radial_distribution}).

\subsection{V-band absolute magnitude}
\label{subsec: abs.magnitude}
Finally, we estimate the absolute magnitude of each stellar system in a similar manner to that described in \citet{Martin2008}. 
In summary, we extract stars randomly one by one and sum up their V-band luminosities from the stellar population model \citep{Bressan2012} with an age of 13~Gyr and metallicity of [M$/$H]$=-2.2$ until the number of stars brighter than the threshold reaches $N_{\ast}$, and it is our magnitude limit ($i=24.5$ mag) converted to the absolute magnitude $V$ using the formula given in \citet{Jordi2006} and the distance modulus $(m-M)_0$. In this way, we perform a Monte Carlo procedure of 10,000 cases and estimate the most likely value of $M_V$ and its uncertainty including that of $N_{\ast}$ and $(m-M)_0$. Here we choose three initial mass functions (IMFs) for the stellar population: from \citet{Kroupa2002}, \citet{Salpeter1955}, and \citet{Chabrier2001}. Unlike total stellar mass, the absolute magnitudes of stellar systems are quite insensitive to the IMFs, because the number of low mass stars has little effect on the total luminosity (Table \ref{tab:4}). We note that these $M_{{\rm tot},V}$ for Vir~I, Cet~III, and Boo~IV are somewhat brighter than what we derived in our previous work \citep{Homma2018,Homma2019}, where the V-band absolute magnitudes for stars brighter than our magnitude limit ($i=24.5$~mag) were presented.

With this method, for Sext~II and Vir~III, we obtain the V-band absolute magnitudes of $-3.91^{+0.37}_{-0.36}$~mag and $-2.69^{+0.45}_{-0.56}$~mag, respectively.

\begin{table*}
\tbl{Estimation of $M_{{\rm tot},V}$ and $M_{\ast}$ from various IMFs}{
\begin{tabular}{l|ccc|ccc}
\hline
Name    &    \multicolumn{3}{|c|}{$M_{{\rm tot},V}$}    &    \multicolumn{3}{|c}{$M_{\ast}$}    \\
    &    \multicolumn{3}{|c|}{[mag]}    &    \multicolumn{3}{|c}{[$M_{\odot}$]}    \\
    &    Kroupa    &    Salpeter    &    Chabrier    &    Kroupa    &    Salpeter    &    Chabrier    \\
\hline
Leo IV    &    $-4.50^{+0.35}_{-0.31}$    &    $-4.52^{+0.32}_{-0.33}$    &    $-4.49^{+0.37}_{-0.28}$    &    $7.4^{+2.4}_{-1.9}\times10^3$    &    $1.4^{+0.5}_{-0.4}\times10^4$    &    $7.6^{+1.7}_{-2.3}\times10^3$    \\
Leo V    &    $-3.96^{+0.35}_{-0.34}$    &    $-4.05^{+0.38}_{-0.30}$    &    $-3.97^{+0.35}_{-0.33}$    &    $4.7^{+1.5}_{-1.2}\times10^3$    &    $9.1^{+2.9}_{-2.4}\times10^3$    &    $4.5^{+1.4}_{-1.2}\times10^3$    \\
Pegasus III    &    $-4.11^{+0.42}_{-0.25}$       &    $-4.08^{+0.35}_{-0.31}$    &    $-4.07^{+0.41}_{-0.26}$    &    $5.1\pm1.4\times10^3$    &    $9.9^{+2.4}_{-3.0}\times10^3$    &    $4.8\pm1.4\times10^3$    \\
Virgo I    &    $-0.90^{+0.72}_{-0.65}$    &    $-0.83^{+0.57}_{-0.74}$    &    $-0.84^{+0.65}_{-0.73}$    &    $3.1^{+2.0}_{-1.0}\times10^2$    &    $6.7^{+3.0}_{-2.5}\times10^2$    &    $3.2^{+1.6}_{-1.2}\times10^2$    \\
Cetus III    &    $-3.45^{+0.49}_{-0.43}$    &    $-3.43^{+0.42}_{-0.48}$    &    $-3.36^{+0.42}_{-0.49}$    &    $2.6^{+1.4}_{-1.0}\times10^3$    &    $5.3^{+2.5}_{-2.1}\times10^3$    &    $2.4^{+1.4}_{-0.9}\times10^3$    \\
Bootes IV    &    $-5.34^{+0.29}_{-0.17}$    &    $-5.38^{+0.28}_{-0.18}$    &    $-5.30^{+0.26}_{-0.20}$    &    $1.6^{+0.4}_{-0.3}\times10^4$    &    $3.0^{+0.8}_{-0.5}\times10^4$    &    $1.6^{+0.3}_{-0.4}\times10^4$    \\
Sextans II    &    $-3.91^{+0.37}_{-0.36}$    &    $-3.98^{+0.40}_{-0.33}$    &    $-3.88^{+0.37}_{-0.36}$    &    $4.3^{+1.5}_{-1.1}\times10^3$    &    $8.4^{+2.8}_{-2.3}\times10^3$    &    $3.9^{+1.6}_{-0.8}\times10^3$    \\
Virgo III    &    $-2.69^{+0.45}_{-0.56}$    &    $-2.81^{+0.49}_{-0.47}$    &    $-2.70^{+0.48}_{-0.53}$    &    $1.5^{+0.7}_{-0.5}\times10^3$    &    $2.9^{+1.4}_{-0.9}\times10^3$    &    $1.4^{+0.7}_{-0.5}\times10^3$    \\
\hline
\end{tabular}}\label{tab:4}
\begin{tabnote}

\end{tabnote}
\end{table*}

\section{Discussion}

\subsection{Comparison with globular clusters and other MW satellites}
In order to determine whether both Sext II and Vir III are in fact MW UFDs, their structural properties are compared with those of MW globular clusters and known dwarf satellites. Globular clusters are known to be self-gravitating stellar systems having compact sizes and round shapes, whereas dwarf satellites are characterized by a diffuse, extended distribution of stars having an ellipitical shape. Also, in contrast to globular clusters, the line-of-sight velocity dispersions of members stars in dwarf satellites suggest that they are surrounded by a massive dark halo.

\begin{figure}[t!]
\begin{center}
\includegraphics[width=85mm]{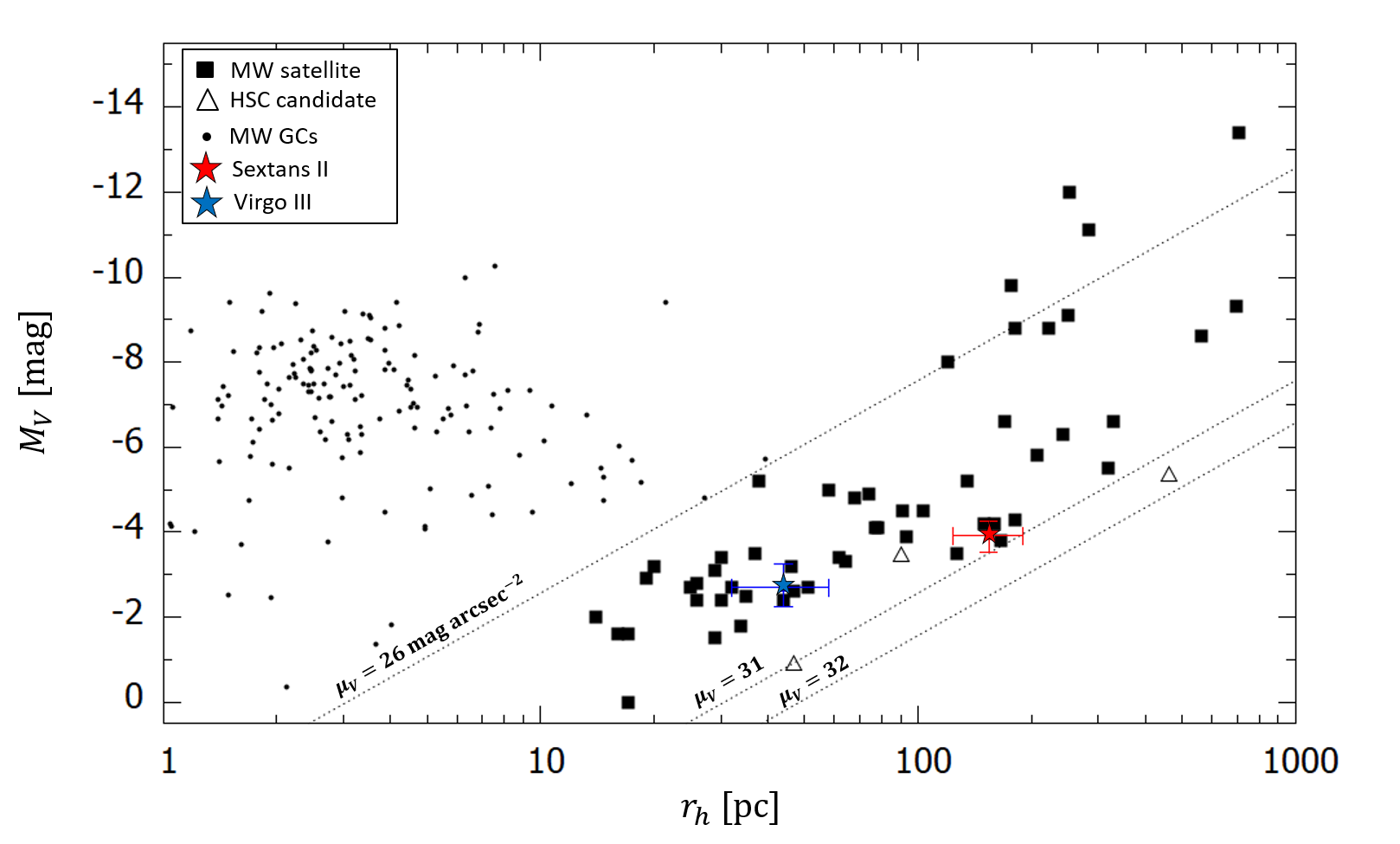}
\end{center}
\caption{The relation between $M_V$ (mag) and $r_h$ (pc) for stellar systems. Dots denote globular clusters in the MW taken from \citet{Harris1996}. Filled squares denote the MW dSphs taken from \citet{McConnachie2012}, the recent DES work for new ultra-faint MW dSphs \citep{Bechtol2015,Koposov2015,Drlica-Wagner2015}, and other recent discoveries \citep{Laevens2014,Kim2015,KimJerjen2015, Laevens2015a,Laevens2015b,Cerny2023}, whereas open triangles denote those found from the
previous data release of the HSC-SSP \citep{Homma2016,Homma2018,Homma2019}. The red and blue stars with error bars correspond to Sextans II and Virgo III, respectively. The dotted lines are lines of the constant surface brightness, $\mu_V = 26$, 31, and 32 mag~arcsec$^{-2}$.
}
\label{fig: abs_size}
\end{figure}

Figure \ref{fig: abs_size} shows the relation between the half-light radii, $r_h$, and V-band absolute magnitudes, $M_V$, for MW dwarf satellites (squares) \citep{McConnachie2012, Laevens2014, Bechtol2015, Koposov2015, Drlica-Wagner2015, Kim2015, KimJerjen2015, Laevens2015a, Laevens2015b, Torrealba2016, Torrealba2018, Cerny2023} and globular clusters (dots) \citep{Harris1996}. Sext II and Vir III are depicted as the red and blue stars with error bars, respectively. This figure indicates that these stellar systems have systematically larger $r_h$ than those of MW globular clusters having similar $M_V$, and that these are located inside the locus of MW dwarf satellites. Also, their shapes are quite flattened ($\varepsilon = 0.3 \sim 0.4$), like other dwarf satellites, and much more so than globular clusters \citep{Harris1996, Marchi2019}.
Thus, we conclude that Sext II and Vir III are new UFDs in the MW.

\begin{figure}[t!]
\begin{center}
\includegraphics[width=85mm]{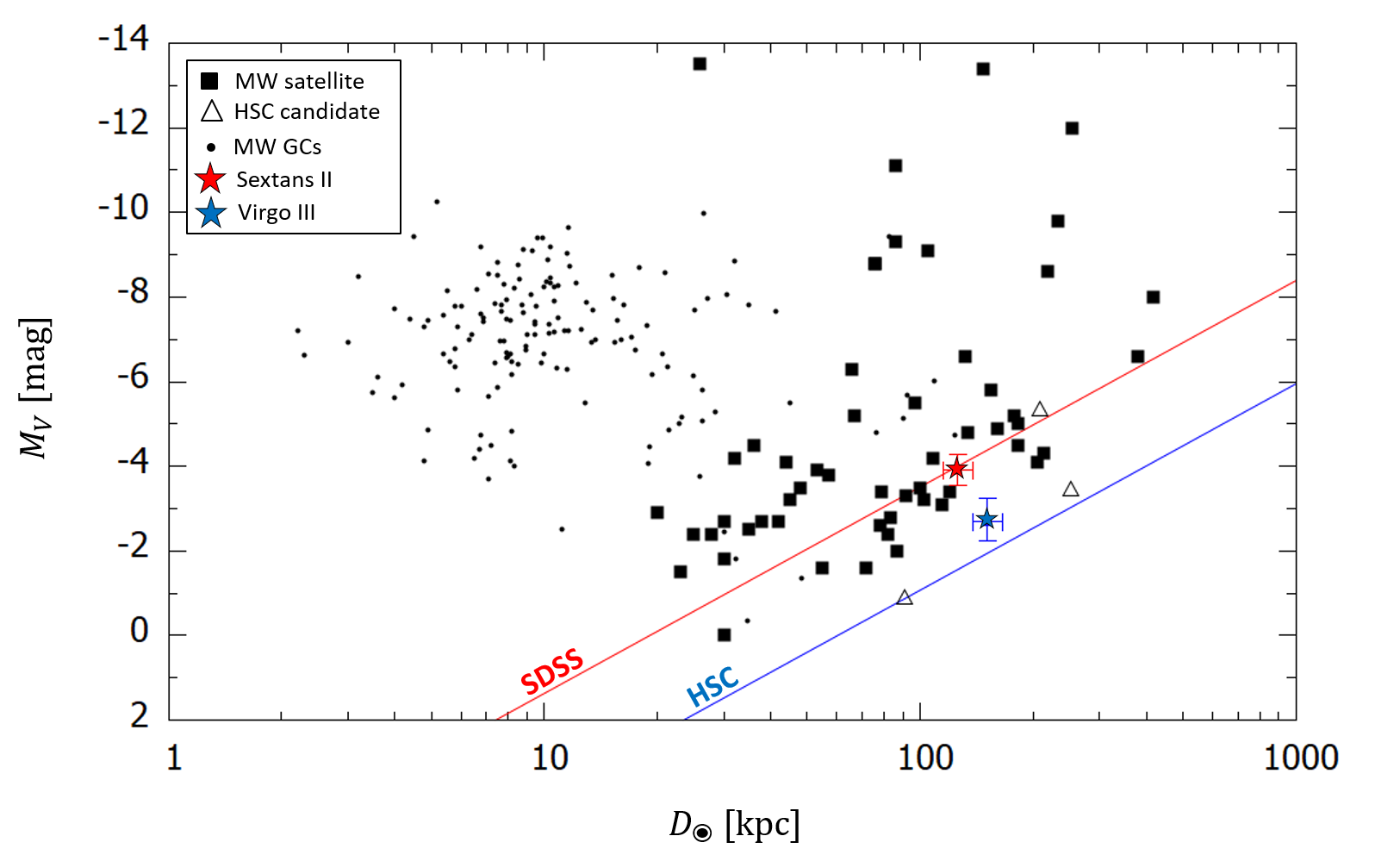}
\end{center}
\caption{The relation between $M_V$ (mag) and heliocentric distance (kpc) for the systems shown in Figure~\ref{fig: abs_size}. The red and blue lines indicate the detection limits of SDSS and HSC, respectively.
}
\label{fig: abs_distance}
\end{figure}

Next, we compare the heliocentric distances of Sext II and Vir III with the distance distributions of MW globular clusters and dwarf satellites. Figure~\ref{fig: abs_distance} shows the relation between $M_V$ and $D_{\odot}$ for these stellar systems, where red and blue stars with error bars correspond to Sext II and Vir III, respectively. In this diagram, red and blue lines show the detection limits of SDSS and HSC, respectively. The former limit is taken from the SDSS completeness distance given as $R_{\rm comp}^{\rm SDSS} = 10^{(-a^\ast M_V - b^\ast)}$~Mpc with $(a^\ast, b^\ast) = (0.205, 1.72)$ \citep{Koposov2008}. The latter limit for HSC, $R_{\rm comp}^{\rm HSC}$, is evaluated from $R_{\rm comp}^{\rm HSC} / R_{\rm comp}^{\rm SDSS} = 10^{0.2(m_{r,{\rm HSC}}-m_{r,{\rm SDSS}})}$, where $m_{r,{\rm SDSS}}$ and $m_{r,{\rm HSC}}$ are the limiting point-source magnitudes in $r$-band for these surveys, given as $m_{r,{\rm SDSS}} = 22.2$~mag and $m_{r,{\rm HSC}} = 24.7$~mag, respectively \citep{Tollerud2008}. The latter value of $m_{r,{\rm HSC}}$ is taken from our cut-off $i$-band magnitude of $i=24.5$~mag and typical $r-i$ colors of stars in MW satellites of $r-i = 0.2$~mag \citep{Homma2019}. As is clear from this figure, both Sext II and Vir III are located at heliocentric distances beyond the radial distribution of typical globular clusters and are also beyond the detection limit of SDSS.

\subsection{Implication for $\Lambda$ CDM models}

In this S21A data release of the HSC-SSP survey, two new satellites, Sext II and Vir III, have been discovered, in addition to the re-identification of three satellites, Vir I, Cet III and Boo IV, which were reported earlier using the previously released HSC-SSP data \citep{Homma2016, Homma2018, Homma2019}. As we have presented in this work, there are four more known satellites in the HSC footprint, Sextans, Leo IV, Leo V and Peg III, indicating the presence of nine satellites in total over the area of $\sim 1,200$~deg$^2$. The implication of this discovery rate of satellites for $\Lambda$CDM models is discussed here, by adopting the latest studies of the expected population of satellites in $\Lambda$CDM models by \citet{Nadler2020} (see also \cite{Nadler2019, Drlica-Wagner2020, Dooley2017, Newton2018}).

To obtain the underlying MW satellite population, \citet{Nadler2020} adopt a high-resolution cosmological zoom-in simulation for MW-mass host halos selected from the work of Mao et al. (2015). In particular, they select two MW-like halos, which are required to have an LMC analog companion galaxy, i.e., having a similar mass, heliocentric distance, and accretion time to those of LMC, to include the impact of the LMC in the analysis. They then consider a detailed model of the galaxy-halo connection, to associate satellite galaxies with subhalos in the simulations \citep{Nadler2019}, using abundance matching, and taking into account satellite sizes, satellite disruption due to baryonic effects, galaxy formation efficiency, and orphan satellites. This model is then marginalized over these astrophysical uncertainties in the fit, to the observed satellite populations by DES and PS1 surveys over the very large area of $\sim 15,000$~deg$^2$. Based on this model, they arrive at a total number of $N_{\rm tot} \equiv N(M_V < 0) = 220 \pm 50$ satellites with $r_h > 10$~pc within the virial radius of the MW. In what follows, we apply this model to our results using the HSC-SSP Wide layer. 

First, we make a correction for $N(M_V)$ associated with the survey area of HSC-SSP over $1,141$~deg$^2$, under the assumption of an isotropic distribution of satellites. This area corresponds to a sky fraction of $f_{\Omega,{\rm HSC}} = 0.028$. Thus, the number of predicted satellites over this sky area is $f_{\Omega,{\rm HSC}} N(M_V)$.

\begin{figure}[t!]
\begin{center}
\includegraphics[width=85mm]{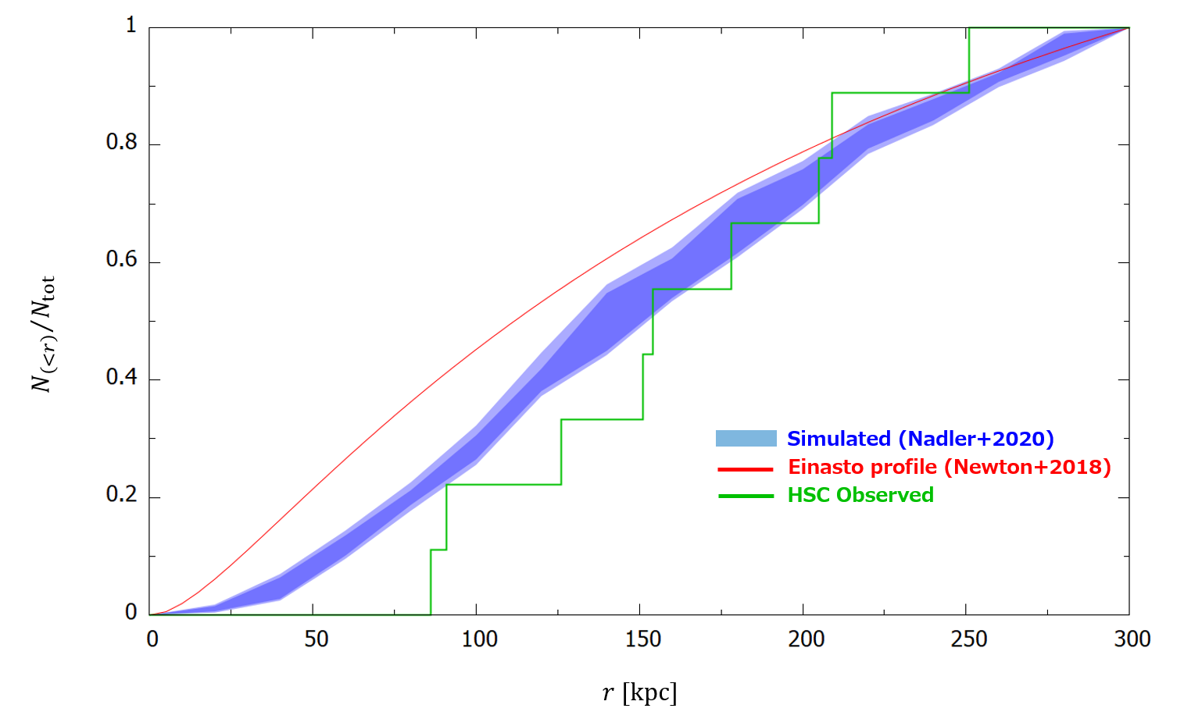}
\end{center}
\caption{Expected radial distribution of MW satellites reproduced from the work of \citet{Nadler2020} (blue bands), compared to the observed satellite distribution from HSC (green histogram). The red curve shows the Einasto profile fit to the radial distribution of satellites predicted by \citet{Newton2018}.
}
\label{fig: radial_dist}
\end{figure}

Second, we make a completeness correction associated with the detection limit of the HSC-SSP survey, which can be expressed in terms of the completeness distance, $R_{\rm comp}^{\rm HSC}(M_V)$, beyond which a satellite with particular $M_V$ cannot be detected (Figure~\ref{fig: abs_distance}).  We then evaluate the completeness correction factor, $f_{\rm r, HSC}(M_V) = N(<R_{\rm comp}^{\rm HSC}(M_V)) / N_{\rm tot}$, following the work of \citet{Nadler2020} to obtain the expected radial distribution of satellites using their code available at GitHub\footnote{\url{github.com/eonadler/subhalo_satellite_connection}. We thank Ethan Nadler for allowing us to use this code for this paper and for his advice in reproducing the results.}. The result is shown in Figure~\ref{fig: radial_dist} (blue bands), from which we evaluate $N(<r)$.
The total number of satellites expected for the HSC-SSP data is then given as $f_{r,{\rm HSC}}f_{\Omega,{\rm HSC}} N(M_V)$.

\begin{figure}[t!]
\begin{center}
\includegraphics[width=85mm]{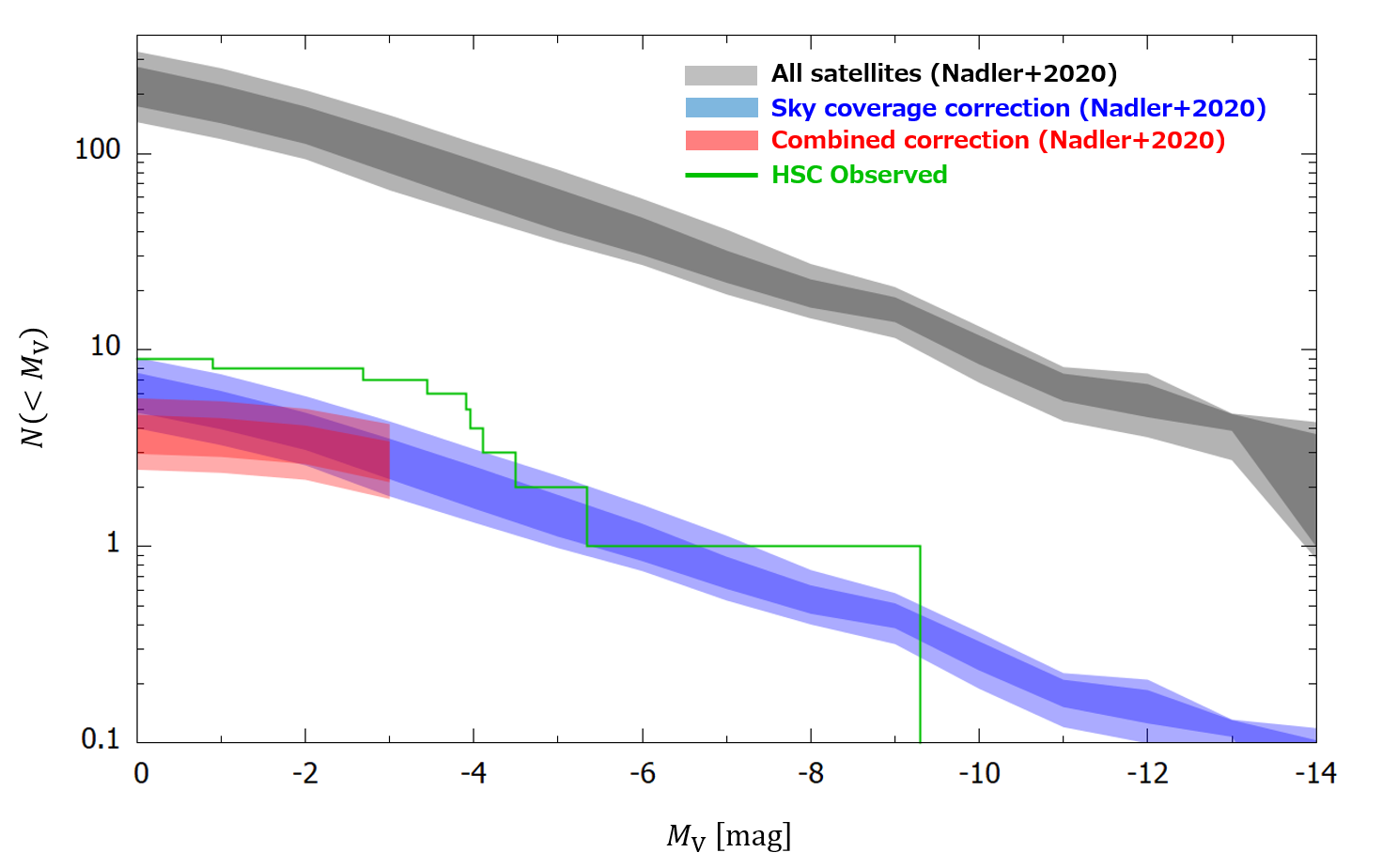}
\end{center}
\caption{
Expected number of MW satellites obtained from the work of \citet{Nadler2020} ($N(M_V)$, black bands), sky-coverage corrected count ($f_{\Omega,{\rm HSC}} N(M_V)$, blue bands), and corrected further for the magnitude limit ($f_{r,{\rm HSC}}f_{\Omega,{\rm HSC}} N(M_V)$, red bands), where dark and light bands correspond to 68\% and 95\% confidence intervals, respectively. Green histogram shows the observed luminosity function obtained from the current work with HSC except Sextans($M_V=-9.3$~mag, \cite{McConnachie2012}).
}
\label{fig: lumi_fun}
\end{figure}

Figure~\ref{fig: lumi_fun} shows the counts for all satellites ($N(M_V)$, black bands), for the sky-coverage correction ($f_{\Omega,{\rm HSC}} N(M_V)$, blue bands), and for the combined correction ($f_{r,{\rm HSC}}f_{\Omega,{\rm HSC}} N(M_V)$, red bands), where dark and light bands correspond to 68\% and 95\% confidence intervals, respectively. For $M_V < 0$ (and $r_h>10$~pc, which is fulfilled for the observed satellites), the total number of satellites expected for the HSC-SSP footprint, $f_{r,{\rm HSC}}f_{\Omega,{\rm HSC}} N(M_V)$, is estimated as $3.9 \pm 0.9$. This expected number is significantly smaller than the nine satellites actually observed, whose cumulative luminosity distribution is shown as the green solid line in Figure~\ref{fig: lumi_fun}. Thus, instead of a missing satellites problem, this is a problem of too many satellites in the HSC-SSP footprint. We had come to a similar conclusion with the S18A data release \citep{Homma2019}. 

This tension is not solved even if we adopt a different model for the radial distribution of satellites. For instance, we adopt the Einasto profile,
\begin{equation}
    \ln N(r)/N_{-2} = - (2/\alpha) [(r/r_{-2})^\alpha - 1] \ ,
\end{equation}
as shown in the model of \citet{Newton2018}, where $\alpha$ is a shape parameter, $r_{-2}$ is the radius at which the logarithmic slope is $-2$, and $N_{-2}$ is the density at $r=r_{-2}$. The concentration of this density profile is parameterized by $c_{200} = R_{200}/r_{-2}$, where $R_{200}$ is the virial radius. If we adopt $\alpha=0.24$, $c_{200}=4.0$, and $R_{200}=200$~kpc (derived for a $1.0\times10^{12} M_{\odot}$ MW-mass halo \citep{Newton2018}), we obtain the expected number of satellites in the HSC footprint as $4.3 \pm 1.0$, which again falls short of the nine observed satellites.

While this high detection probability of satellites in the HSC-SSP appears to solve the missing satellites problem in $\Lambda$CDM models, the remained tension with the predicted models for the population of MW satellites needs to be settled. A possible key is the observed radial distribution of nine satellites in the HSC footprint as shown in the green line in Figure~\ref{fig: radial_dist}. Compared with the model result (blue bands), the observed radial distribution is more extended, although we are limited by small-number statistics. However, taking into account that the Subaru/HSC deep imaging survey is more sensitive to faint, distant MW satellites than other existing surveys, it is possible that the true radial distribution of MW satellites is more extended than that derived from the DES and PS1 surveys, so that the matching of the simulated subhalos with the observed MW satellites needs to be reconsidered. Indeed, in this more extended case, the predicted number of satellites could be larger than the case of the currently adopted model by \citet{Nadler2020}, settling the tension with HSC-SSP survey counts \citep{Kim2018}. Also, related to this point, more numerical simulation data for MW-sized dark halos having an LMC analog are needed for getting the higher statistical significance in these results.

Secondly, it should also be borne in mind that the detectability of satellites in the HSC-SSP is simplified by assuming it to be a step function which changes from $100\%$ to $0\%$ beyond a particular radius calculated as the detection limit distance ($R_{\rm comp}$) at each magnitude. Note that the detection limit from \citet{Koposov2008}, which we adopt for our completeness boundary, indicates the $50\%$ detection rate only to objects with surface brightness $\mu_V \leq30$~mag~arcsec$^{-2}$ and ignores the complicated issue that the detectability varies with a location in the surveyed area due to the crowdness of the sky \citep{Dolinsky2022, Qu2023}. Given this fact, in reality there should be a greater dispersion in the expected number of MW satellites in the HSC-SSP than the red band in Figure~\ref{fig: lumi_fun}, and the discrepancy between simulations and observations could be smaller.

Lastly, yet uncertain assumption is the isotropy of the satellite distribution; bright satellites are aligned in a plane nearly perpendicular to the Galactic plane (e.g., \cite{Kroupa2005,McConnachie2006,
Pawlowski2012,Pawlowski2013,Pawlowski2015}). Hence, the sky-coverage correction (blue bands in Figure~\ref{fig: lumi_fun}) would have fluctuations in the small footprint of HSC-SSP ($2.8\%$ of the whole sky).

In the near future, it is expected that the Vera C. Rubin Observatory LSST \citep{Ivezic2019} over the entire southern sky and the Chinese Space Station Telescope (CSST) survey \citep{Zhan2011,Cao2018,Gong2019} covering $\sim 17,500$ deg$^2$ will discover a significant majority of the MW satellites plus many dwarf galaxies in and out of the Local Group \citep{Qu2023}. These upcoming surveys determine whether this new, too many satellites problem is real, at much higher statistical significance.

\section{Conclusions}

From the HSC-SSP Wide layer data obtained before 2021 January (final internal data release S21A), covering $\sim 1,140$ deg$^2$ in the sky, we have searched for new MW satellites using deep $g$, $r$, and $i$ band photometry. In our previous papers using the released data before 2018 April over $\sim 676$~deg$^2$ (S18A), we reported the discovery of three candidate satellites, Vir~I, Cet~III, and Boo~IV \citep{Homma2016,Homma2018,Homma2019}. Here in the data of S21A, we have re-identified these satellites and also found two new candidates, Sext~II and Vir~III, at $D_{\odot} = 126$~kpc and $151$~kpc, respectively. Sext~II is characterized by $M_V\simeq-3.9$~mag and $r_h\simeq154$~pc, whereas Vir~III has $M_V \simeq-2.7$~mag and $r_h\simeq 44$~pc. Comparison with $M_V$ vs. $r_h$ relations for MW globular clusters and dwarf satellites indicates that these two new candidates most likely can be identified as UFDs. There are four previously known satellites in the HSC footprint, Sextans, Leo~IV, Leo~V, and Peg~III, giving a total of nine satellites. We have presented the structural parameters of these satellites (except Sextans) using the current HSC-SSP photometry data.

To understand implications for the missing satellites problem in $\Lambda$CDM models, we have adopted the recent work for the expected population of MW satellites by \citet{Nadler2020}. Their work combined a model of the galaxy-halo connection with simulated MW-mass host halos that associate an LMC analog and was marginalized to the observed satellite populations by DES and PS1 over $\sim 15,000$~deg$^2$. The expected total number of satellites in their model is $N(M_V<0) = 220 \pm 50$ satellites with $r_h > 10$~pc within the virial radius of the MW. The application of this model to the HSC-SSP survey after making a sky-coverage and completeness corrections suggests that a total of $3.9 \pm 0.9$ satellites are expected in the HSC footprint, compared to the nine observed satellites. This tension is not solved even if we exclude a classical dwarf, Sextans, in the galaxy count or even if we adopt the prediction of a different model for the underlying population of MW satellites \citep{Newton2018}. Therefore, while the high discovery rate of MW satellites, combined with a likely model of the galaxy-halo connection, appears to suggest that the missing satellites problem for $\Lambda$CDM models is no more serious than previously thought, we encounter a new tension, an apparent too many satellites problem. Meanwhile, it is important to recognize that our results are based on small-number statistics and several assumptions such as the isotropic distribution of satellites and the detection limit distance depending only on their luminosity.

To get further insights into this issue, we require spectroscopic follow-up studies of stars in these candidate satellites to constrain their membership as well as their chemo-dynamical properties. We also need to explore high-resolution imaging of these satellites with space telescopes (HST, Euclid, and more) to clearly make star/galaxy separation for faint sources. A tighter constraint on the MW satellite population will be brought by the LSST survey of the Vera C. Rubin Observatory over the entire southern sky and the CSST survey covering $\sim 17,500$ deg$^2$, providing the number density of satellites and their radial distribution at a much higher statistical significance than the current work. 

\begin{ack}
This work is based on data collected at the Subaru Telescope and retrieved from
the HSC data archive system, which is operated by Subaru Telescope and Astronomy
Data Center at National Astronomical Observatory of Japan.
We acknowledge support in part from MEXT Grant-in-Aid for Scientific Research (No.~JP18H05437, JP21H05448, and JP24K00669 for M.C.,
No.~JP20H01895, JP21K13909, and JP23H04009 for K.H.).

The Hyper Suprime-Cam (HSC) collaboration includes the astronomical communities
of Japan and Taiwan, and Princeton University. The HSC instrumentation and
software were developed by the National Astronomical Observatory of Japan (NAOJ),
the Kavli Institute for the Physics and Mathematics of the Universe (Kavli IPMU),
the University of Tokyo, the High Energy Accelerator Research Organization (KEK),
the Academia Sinica Institute for Astronomy and Astrophysics in Taiwan (ASIAA),
and Princeton University. Funding was contributed by the FIRST program from Japanese
Cabinet Office, the Ministry of Education, Culture, Sports, Science and Technology (MEXT),
the Japan Society for the Promotion of Science (JSPS), Japan Science and
Technology Agency (JST), the Toray Science Foundation, NAOJ, Kavli IPMU, KEK, ASIAA,
and Princeton University.  

This paper makes use of software developed for the Large Synoptic Survey Telescope.
We thank the LSST Project for making their code available
as free software at \url{http://dm.lsst.org}.

The Pan-STARRS1 Surveys (PS1) have been made possible through contributions of
the Institute for Astronomy, the University of Hawaii, the Pan-STARRS Project Office,
the Max-Planck Society and its participating institutes, the Max Planck Institute for Astronomy,
Heidelberg and the Max Planck Institute for Extraterrestrial Physics, Garching,
The Johns Hopkins University, Durham University, the University of Edinburgh,
Queen's University Belfast, the Harvard-Smithsonian Center for Astrophysics,
the Las Cumbres Observatory Global Telescope Network Incorporated,
the National Central University of Taiwan, the Space Telescope Science Institute,
the National Aeronautics and Space Administration under Grant No. NNX08AR22G issued through
the Planetary Science Division of the NASA Science Mission Directorate,
the National Science Foundation under Grant No. AST-1238877, the University of Maryland,
and Eotvos Lorand University (ELTE) and the Los Alamos National Laboratory.
\end{ack}


\appendix
\section{Space, CMDs, and density distributions for other seven satellites}
\label{sec: appendix_1}

Here we show the space and CMD distributions for four known satellites, Sextans, Leo~IV, Leo~V, and Peg~III, and candidate UFDs that we reported in our previous papers \citep{Homma2018,Homma2019}, Vir~I, Cet~III, and Boo~IV (Figures 10-16). The density distributions of these satellites are summarized in Figure 17.

\begin{figure*}[h!]
\begin{center}
\includegraphics[width=120mm]{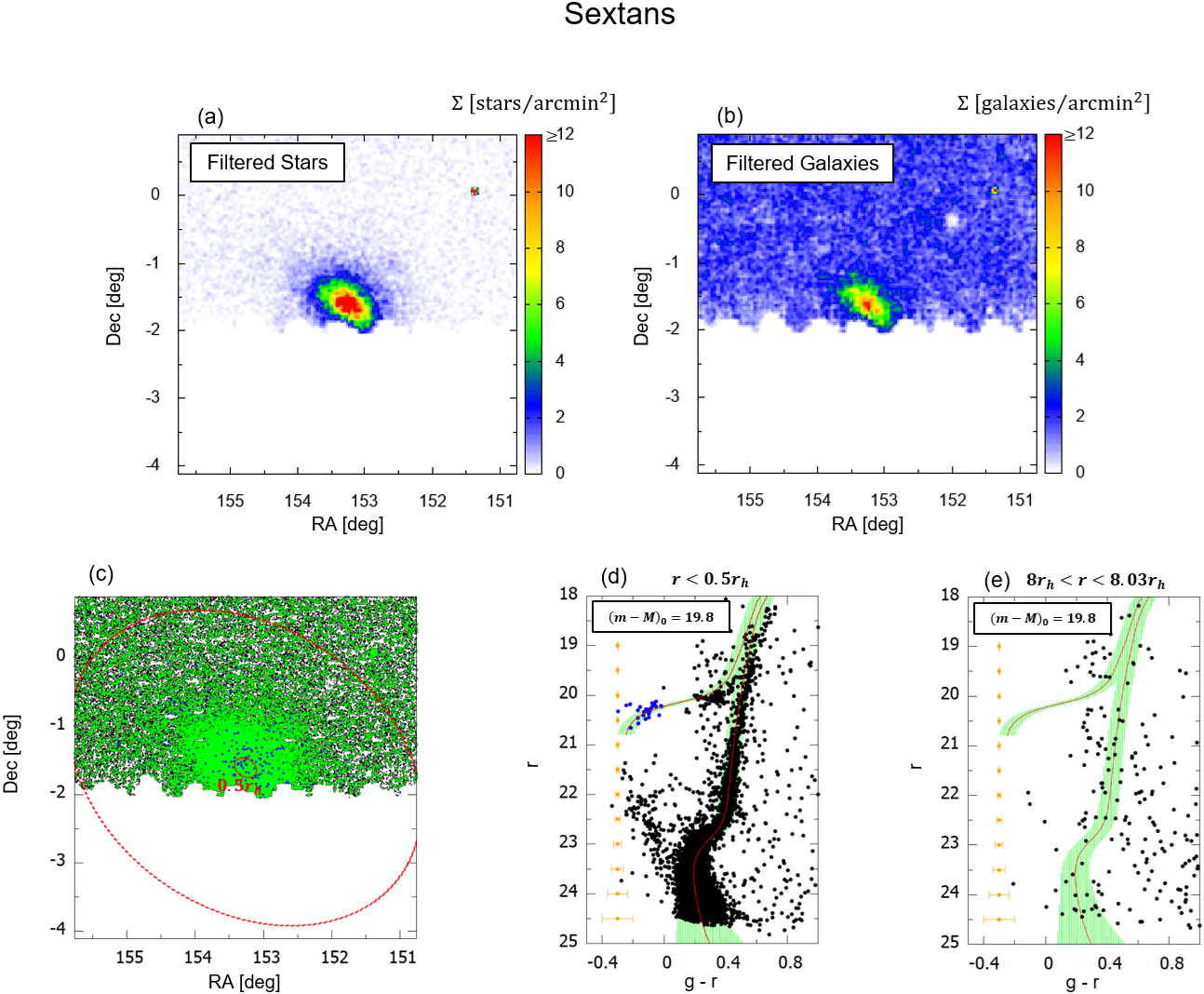}
\end{center}
\caption{
Sextans. Same as Figure~\ref{fig: SextansII_cmd_new} except that the isochrone filter is at $(m-M)_0 = 19.8$.
We note that another overdensity located at (RA, DEC)=(151.3, 0.1)[deg] is the globular cluster, Pal 3 \citep{Harris1996}.
(c) The solid red curve shows an ellipse with a major axis of $r=0.5r_h$($r_h=20.7'$) and an ellipticity of $0.25$ \citep{Tokiwa2023},
whereas dotted red lines show annuli with radii $r=8.00r_h$ and $r=8.03r_h$ from the center of the galaxy. 
}
\label{fig: Sextans_cmd}
\end{figure*}

\begin{figure*}[h!]
\begin{center}
\includegraphics[width=120mm]{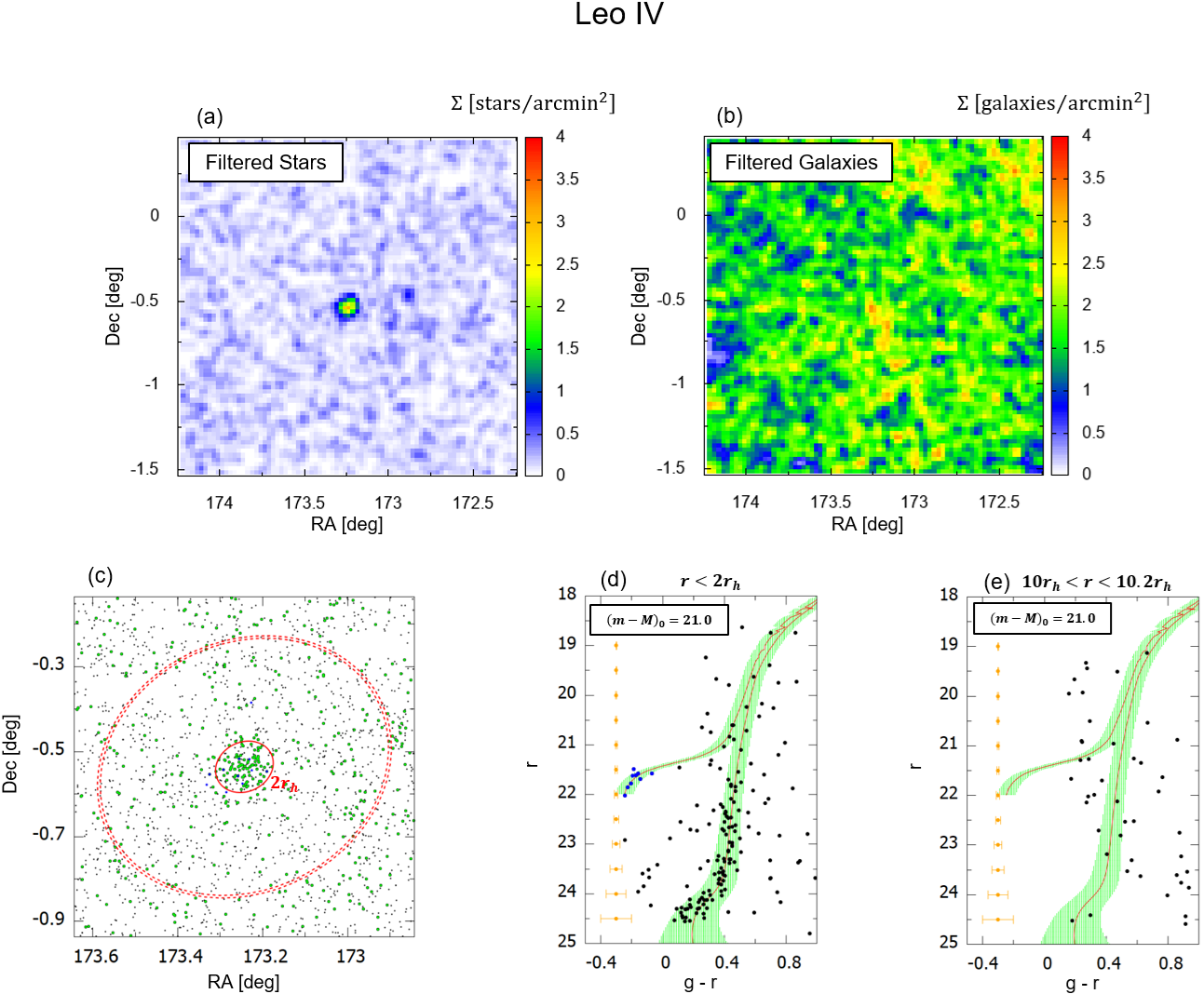}
\end{center}
\caption{
Leo IV. Same as Figure~\ref{fig: SextansII_cmd_new} except that the isochrone filter is at $(m-M)_0 = 21.0$. (c) The solid red curve shows an ellipse with a major axis of $r=2.0r_h$($r_h=2.6'$) and an ellipticity of $0.17$. 
}
\label{fig: LeoIV_cmd}
\end{figure*}

\begin{figure*}[t!]
\begin{center}
\includegraphics[width=120mm]{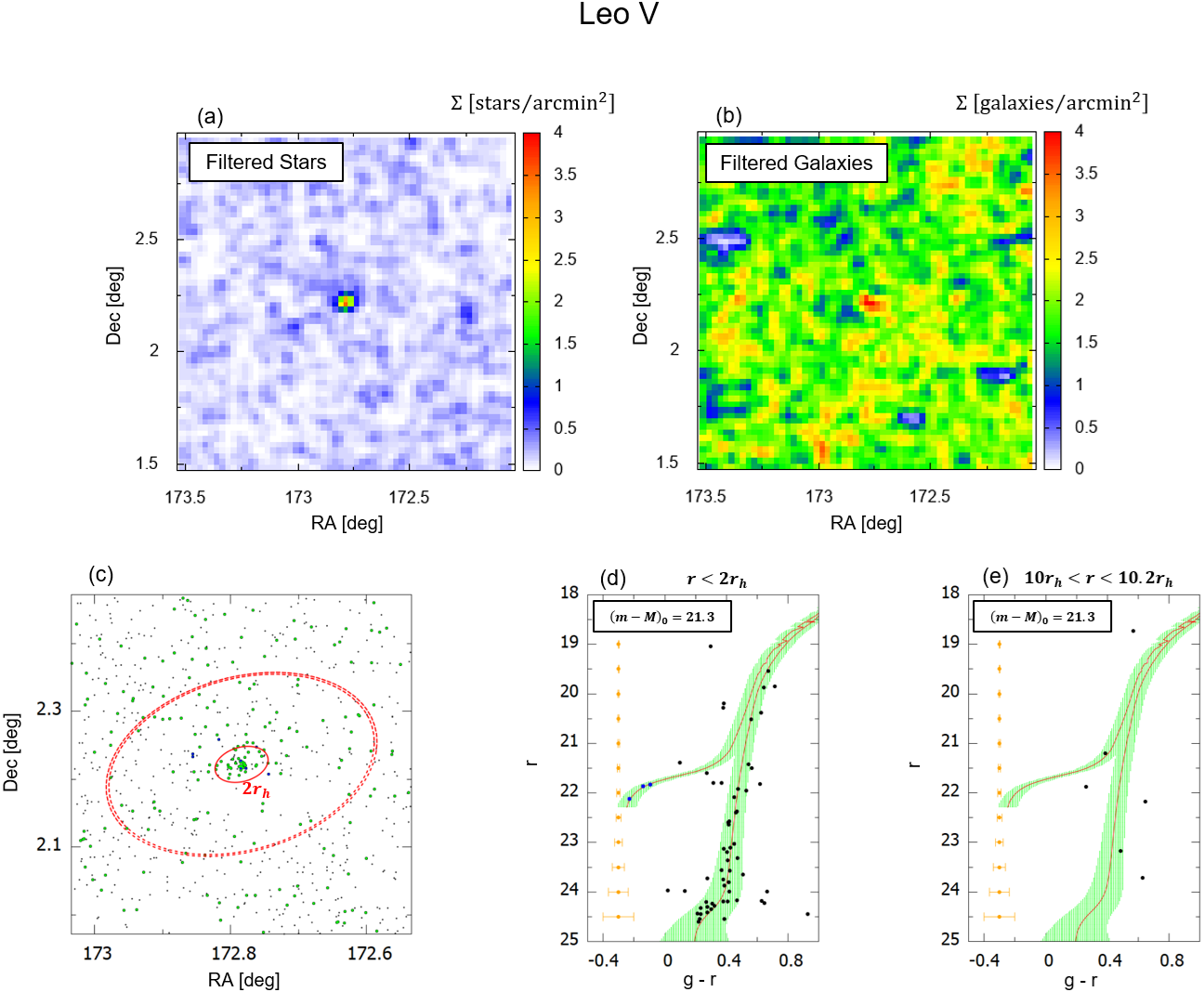}
\end{center}
\caption{
Leo V. Same as Figure~\ref{fig: SextansII_cmd_new} except that the isochrone filter is at $(m-M)_0 = 21.3$. (c) The solid red curve shows an ellipse with a major axis of $r=2.0r_h$($r_h=1.2'$) and an ellipticity of $0.37$. 
}
\label{fig: LeoV_cmd}
\end{figure*}

\begin{figure*}[t!]
\begin{center}
\includegraphics[width=120mm]{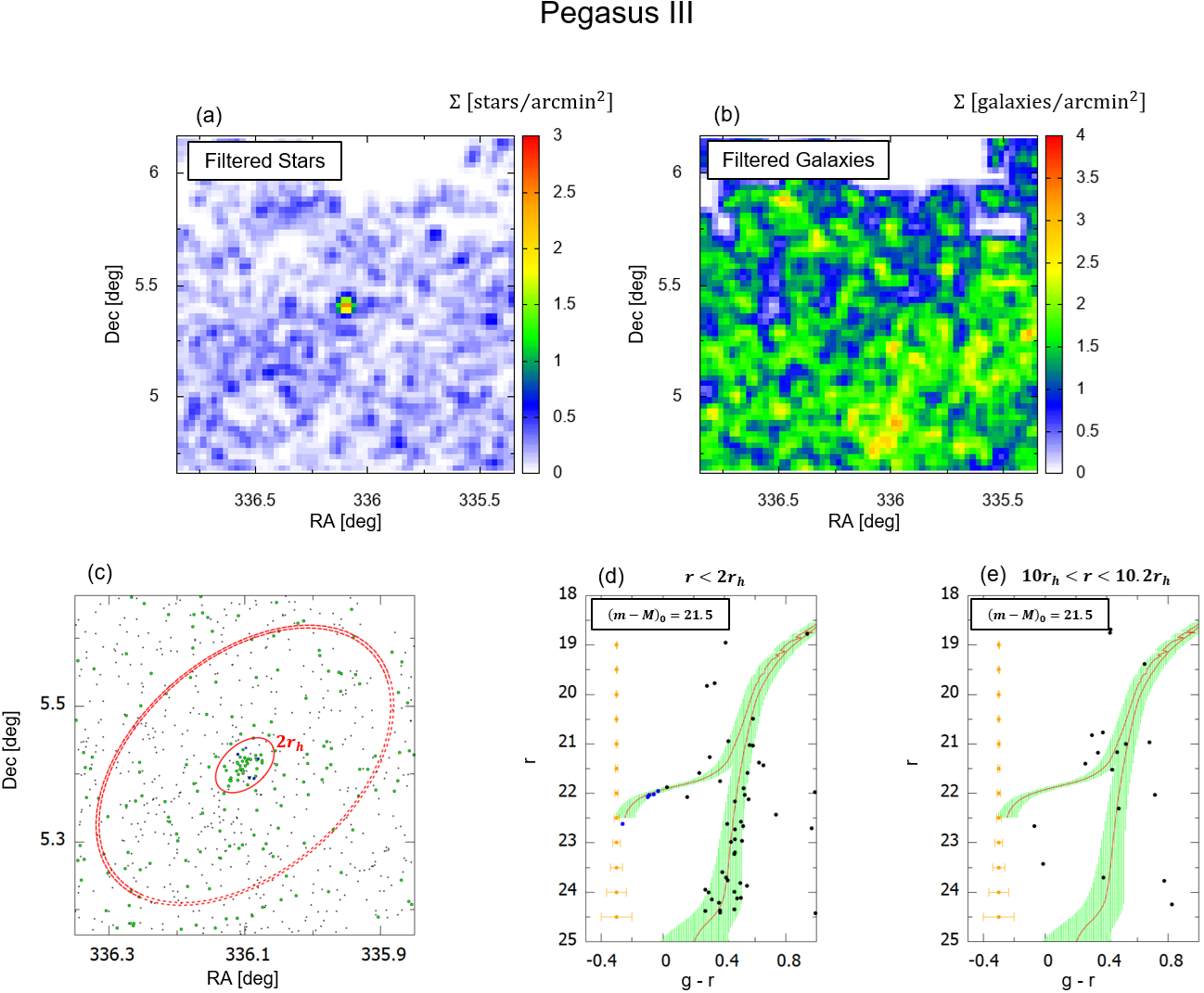}
\end{center}
\caption{
Pegasus III. Same as Figure~\ref{fig: SextansII_cmd_new} except that the isochrone filter is at $(m-M)_0 = 21.5$. (c) The solid red curve shows an ellipse with a major axis of $r=2.0r_h$($r_h=1.5'$) and an ellipticity of $0.37$. 
}
\label{fig: PegasusIII_cmd}
\end{figure*}

\begin{figure*}[t!]
\begin{center}
\includegraphics[width=120mm]{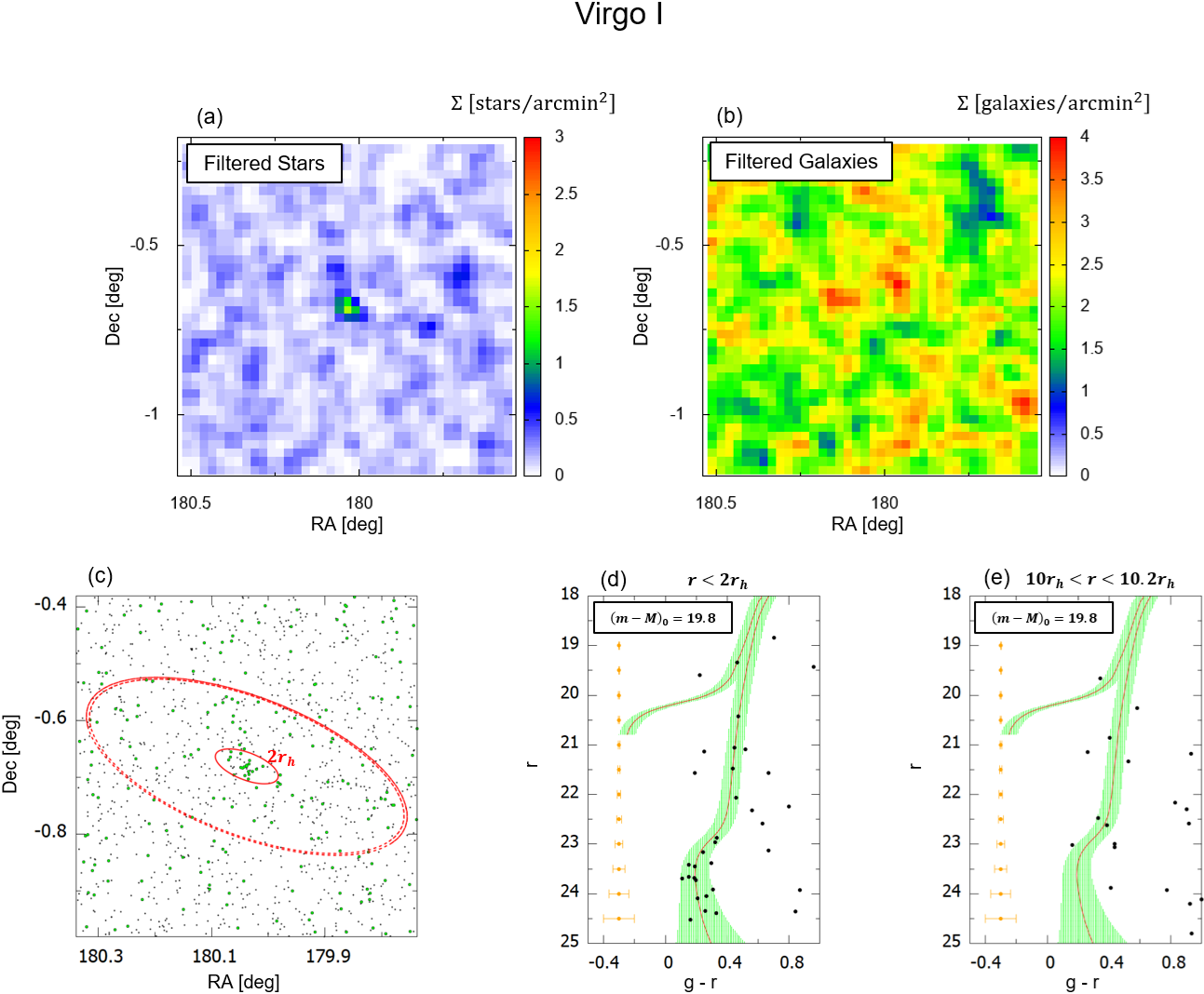}
\end{center}
\caption{
Virgo I. Same as Figure~\ref{fig: SextansII_cmd_new} except that the isochrone filter is at $(m-M)_0 = 19.8$. (c) The solid red curve shows an ellipse with a major axis of $r=2.0r_h$($r_h=1.8'$) and an ellipticity of $0.59$. 
}
\label{fig: VirgoI_cmd}
\end{figure*}

\begin{figure*}[t!]
\begin{center}
\includegraphics[width=120mm]{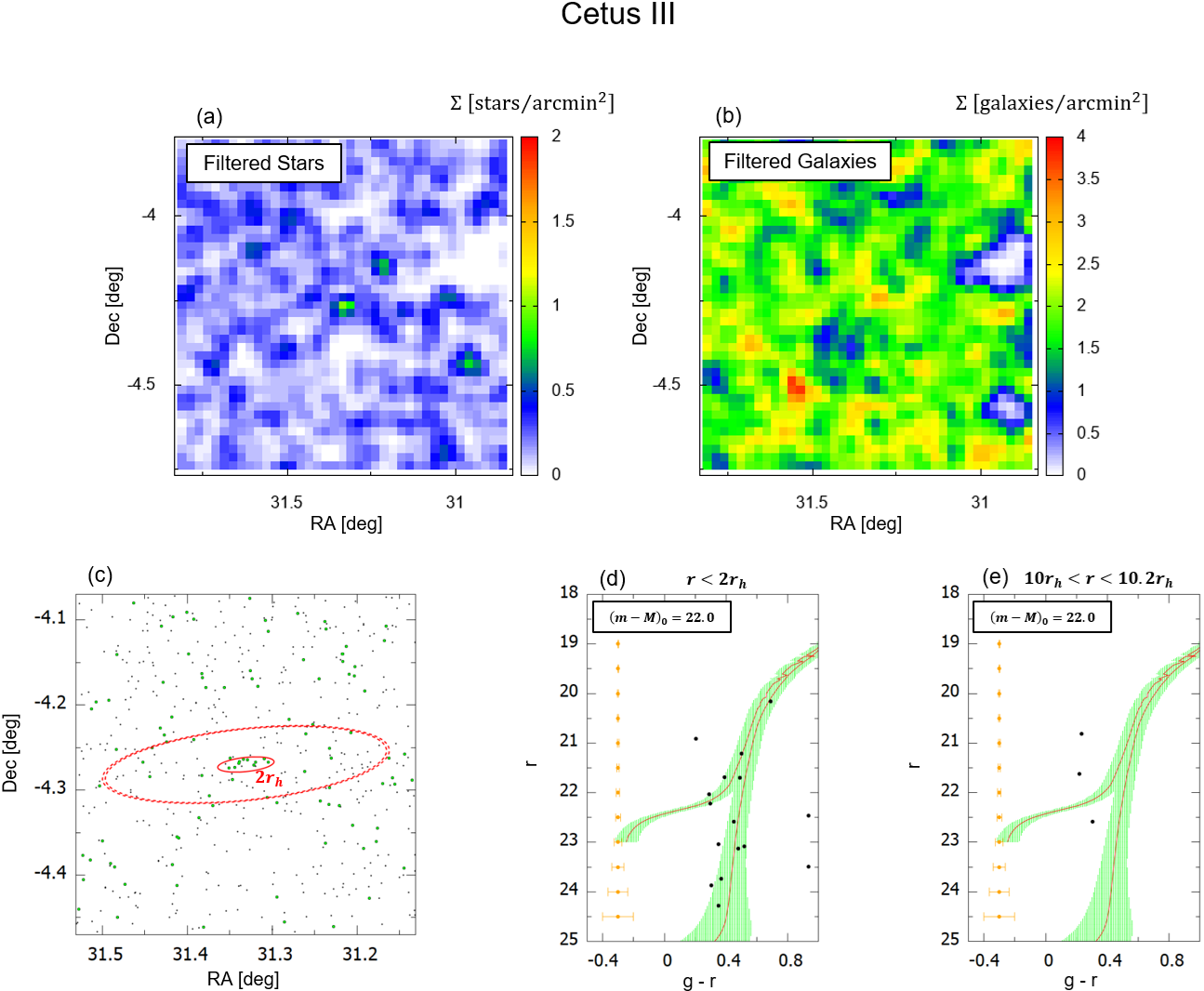}
\end{center}
\caption{
Cetus III. Same as Figure~\ref{fig: SextansII_cmd_new} except that the isochrone filter is at $(m-M)_0 = 22.0$. (c) A solid red curve shows an ellipse with a major axis of $r=2.0r_h$($r_h=1.2$) and an ellipticity of $0.76$. 
}
\label{fig: CetusIII_cmd}
\end{figure*}

\begin{figure*}[t!]
\begin{center}
\includegraphics[width=120mm]{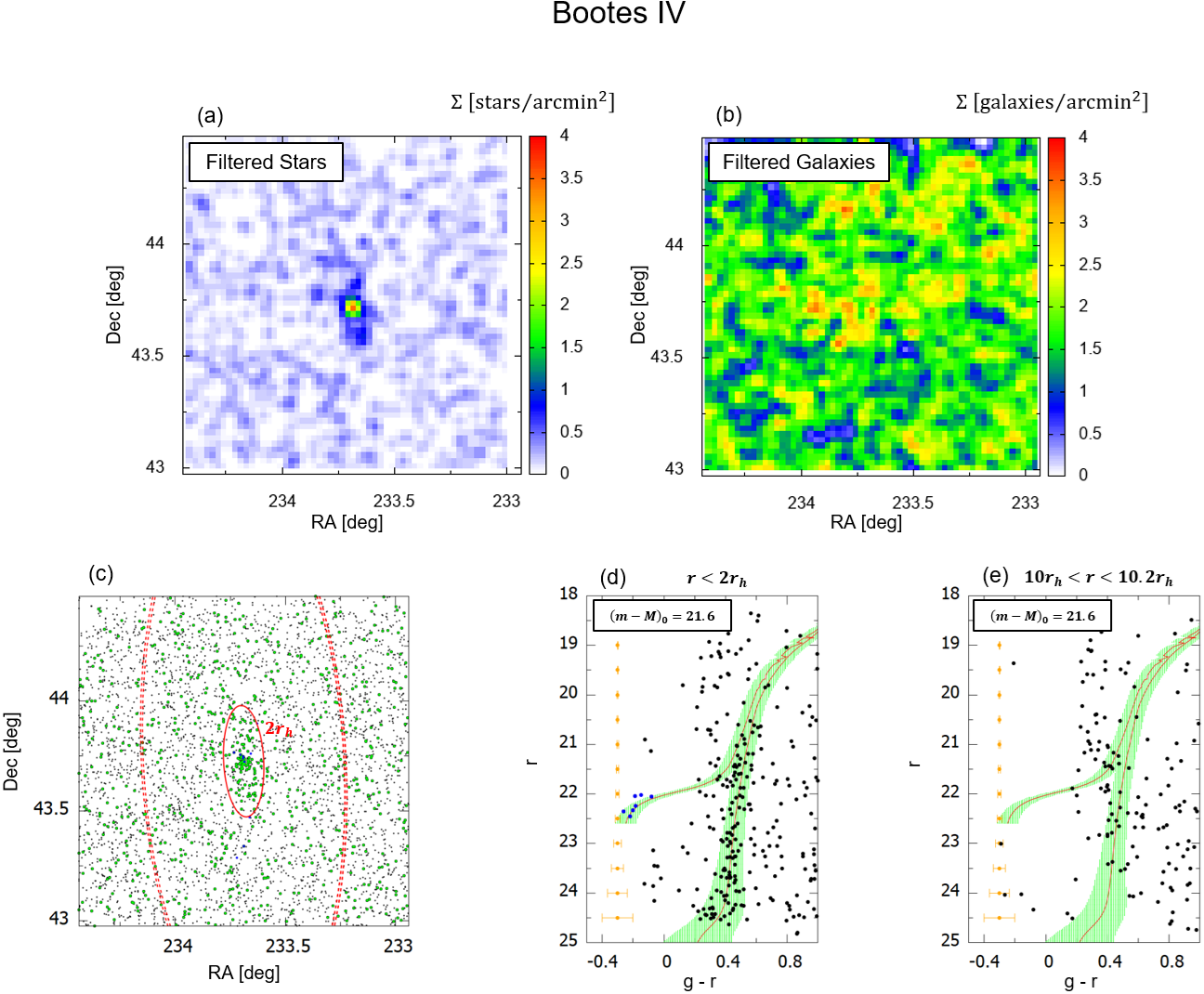}
\end{center}
\caption{
Bootes IV. Same as Figure~\ref{fig: SextansII_cmd_new} except that the isochrone filter is at $(m-M)_0 = 21.6$. (c) A solid red curve shows an ellipse with a major axis of $r=2.0r_h$($r_h=7.6'$) and an ellipticity of $0.64$. 
}
\label{fig: BootesIV_cmd}
\end{figure*}

\begin{figure*}[t!]
\begin{center}
\includegraphics[width=170mm]{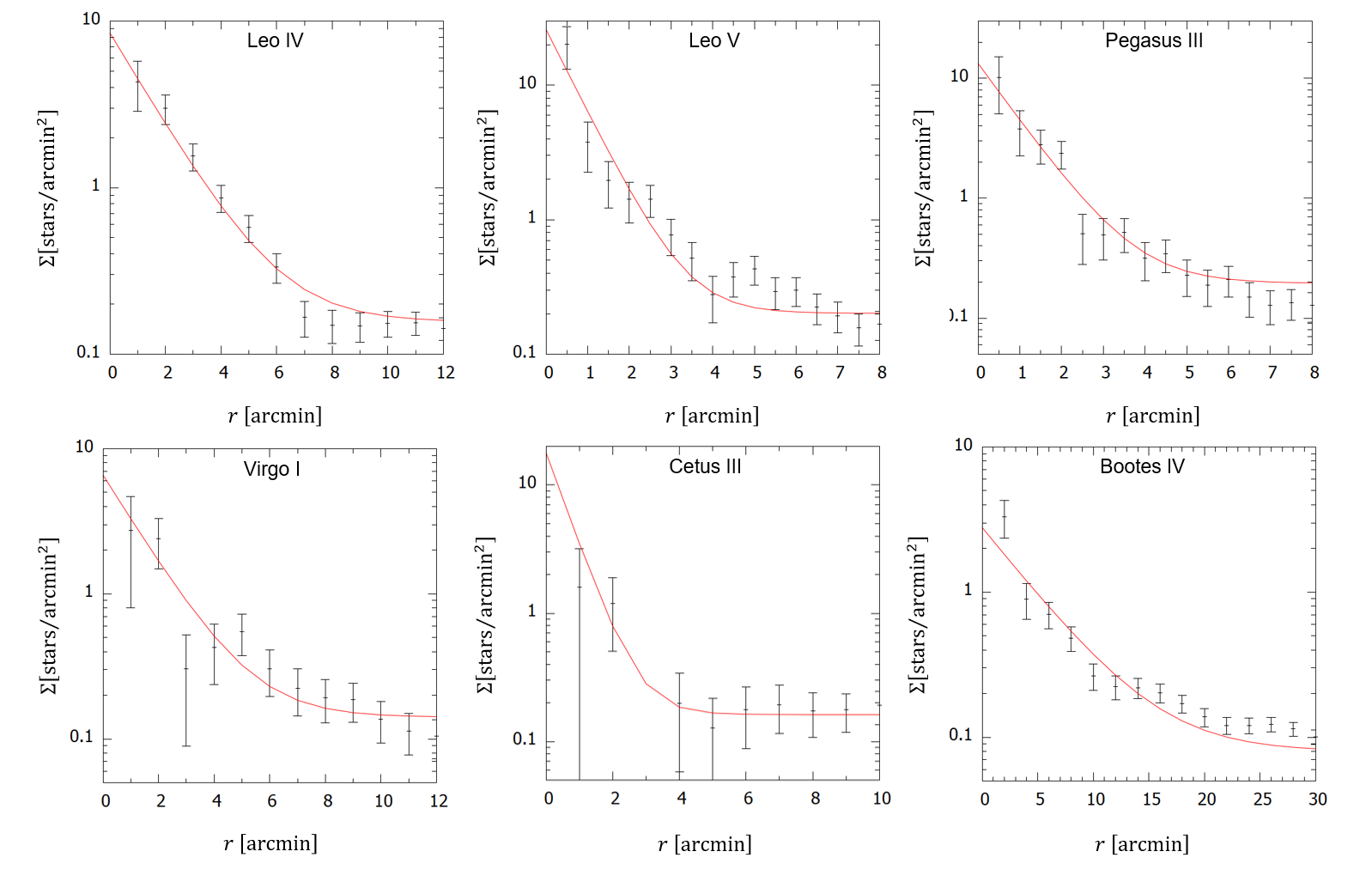}
\end{center}
\caption{
The density distribution of the stars in satellites passing the isochrone filter shown in Figure 10-16(d), in elliptical annuli as a function of mean radius. The uncertainties are based on Poisson statistics. The red line denotes the best-fit exponential profile within $5\sim6r_h$ plus a constant representing the background.

}
\label{fig: Radial_distribution}
\end{figure*}

\end{document}